\documentclass[%
reprint,
 amsmath,amssymb,
 aps,floatfix,
]{revtex4-2}

\usepackage{graphicx}
\usepackage{dcolumn}
\usepackage{bm}
\usepackage{hyperref}
\usepackage{array}

\usepackage{longtable}
\usepackage{booktabs}
\AtBeginEnvironment{quote}{\itshape}
\usepackage{makecell}
\usepackage{tabularx}
\usepackage{ragged2e}
\makeatletter

\usepackage{booktabs}
\usepackage{xcolor}
\usepackage{tabularx}
\usepackage{amssymb}
\usepackage{xspace}
\usepackage{enumitem}

\usepackage[english,provide=*]{babel}
\usepackage{hyperref}

\begin{document}

\newcommand{\insertFigure}[3]{
\begin{figure}[H]
\centering
  \includegraphics[width=\columnwidth]{#1}
  \caption{#3\label{#2}}
\end{figure}
}

\newcommand{\insertSmallFigure}[3]{
\begin{figure}[H]
\centering
  \includegraphics[width=.5\columnwidth]{#1}
  \caption{#3\label{#2}}
\end{figure}
}

\newcommand{\insertFigureFullPage}[3]{
\begin{figure*}[H]
\centering
  \includegraphics[width=0.98\textwidth]{#1}
  \caption{#3\label{#2}}
\end{figure*}
}

\newcommand{\insertFigureVarWidth}[4]{
\begin{figure*}[H]
\centering
  \includegraphics[width=#3]{#1}
  \caption{#4\label{#2}}
\end{figure*}
}

\newcommand{\insertDoubleFigure}[4]{
\begin{figure*}[H]
\centering
  \includegraphics[width=0.49\textwidth]{#1}
  \includegraphics[width=0.49\textwidth]{#2}
  \caption{#4\label{#3}}
\end{figure*}
}

\newcommand{\checkThis}[1]{\textcolor{red}{#1}}
\newcommand{\revChange}[1]{\textcolor{blue}{#1}}
\newcommand{\maple}{MAPLE\xspace}

\title{Design and Initial Evaluation of a Photovoltaics-focused Course-based Undergraduate Research Experience in Physics}

\author{Rachael L. Merritt$^{1,2}$}
  \email{rachael.merritt@colorado.edu, she/her}
\author{H. J. Lewandowski$^{1,2}$}

\affiliation{$^1$Department of Physics, University of Colorado Boulder, Boulder, CO 80309, USA}
\affiliation{$^2$JILA, National Institute of Standards and Technology and the University of Colorado, Boulder, CO 80309, USA }

\date{\today}

\begin{abstract}
Traditional physics laboratory courses often focus on experiments with well-known results, limiting students' engagement in authentic scientific practices. Course-based undergraduate research experiences (CUREs), where students engage in real research with unknown outcomes, have been shown to support positive student outcomes, such as increased self-efficacy, persistence, and engagement in scientific practices. However, discipline-specific studies of CUREs in physics remain limited. We describe the development, structure, and initial implementation of a photovoltaics-focused CURE in a second-year undergraduate physics laboratory course at the University of Colorado Boulder. To examine how students experienced the course, we analyzed end-of-semester reflection assignments using the five CURE components (i.e., scientific practices, discovery, relevance, collaboration, and iteration), as well as established dimensions of research authenticity, as analytic frameworks. Students described experiences associated with all five CURE components, with collaboration, relevance, and scientific practices appearing most prominently in their reflections. Students also associated authentic research with meaningful scientific contribution, engagement in authentic scientific practices, and navigating the uncertainty and setbacks inherent in research, although fewer explicitly identified themselves as researchers or scientists. A subset of students additionally connected the course to their immediate thinking about future academic and professional pathways. This work contributes both a discipline-specific model for implementing CUREs in experimental physics laboratory courses and provides insight into how students interpret and experience authentic research within this course context.
\end{abstract}

\maketitle

\section{Introduction}\label{Intro}
For nearly 150 years, laboratory courses have been an important part of the physics curriculum \cite{may_labhistory2023}. These courses provide students with direct, hands-on opportunities to engage in observation and experimentation. Lab courses have a variety of goals including to help students `think like physicists,' combining theoretical understanding with practical skills, and developing the habits of mind that support scientific inquiry \cite{kozminski2014aapt}. The American Association of Physics Teachers (AAPT) Committee on Laboratories recommended that undergraduate lab courses should support students in developing key skills, including constructing knowledge, modeling, designing experiments, building technical and practical laboratory proficiency, analyzing and visualizing data, and communicating physics effectively \cite{kozminski2014aapt}. 

In studies looking at the landscape of physics laboratory instruction, many lab courses do not give students a choice of which experiments to complete and focus on students completing activities that have results that are already known to the instructor or that confirm results covered in a lecture course \cite{holmeslew2020,geschwindtax_2024}. These courses often consist of prescriptive labs, or activities that provide step-by-step instructions that lead students through a predetermined experiment to verify a known outcome. In this work, we use the term traditional prescriptive labs to describe these verification-style, highly structured activities. 

The highly structured nature of traditional prescriptive labs may shape how students engage with experimental work and what they perceive as the goals of laboratory activities (e.g., \cite{ etkina2010design, deacon2011student, eclass, wieman2015measuring, wilcox2016_openvguided, holmes2017value, holmes2018introductory, hu2017, wilcox2017_skilldev, la2021comparison, walsh2022}. It has been shown that instructional labs that focus on reinforcing understanding of physics concepts do not measurably achieve this goal \cite{wieman2015measuring, holmes2017value} and labs that have prescriptive instructions do not require students to think critically or engage deeply with the underlying physics \cite{holmes2018introductory}. The structured nature of these activities may discourage sense-making and reflection, leading students to complete activities without monitoring or evaluating their reasoning \cite{etkina2010design}. Additionally, some students report they believe that simply following the provided procedure is sufficient for completing lab activities, even when conceptual understanding is an intended goal \cite{hu2017}. When laboratory activities emphasize reproducing known results, students may begin to equate ``successful science” with obtaining expected answers rather than engaging in inquiry or discovery \cite{kretchmer2024} and may resort to questionable research practices to arrive at the expected answer \cite{stein2018}.  Traditional prescriptive labs may also have a negative impact on student motivation, as many students find them uninspiring, potentially reducing interest in continued study of experimental physics \cite{galvez2010}. These findings highlight the importance of designing laboratory experiences that more fully support students’ engagement in authentic experimental practices.

Laboratory activities that incorporate unknown outcomes have been shown to foster a stronger appreciation for the development of scientific knowledge \cite{hu2017} and provide more opportunities for students to engage authentically with experimental physics practices \cite{wilcox2016_openvguided, wilcox2017_skilldev}. One laboratory modality that incorporates open-ended activities and is designed to more directly address the AAPT recommendations is the implementation of course-based undergraduate research experiences (CUREs). CUREs provide students with an authentic research experience as part of a regular course. Unlike traditional research opportunities, which require applications or faculty selection, participation is integrated into the course structure. A defining feature of a CURE is that student work contributes to the generation of new knowledge with relevance beyond the classroom. Auchincloss et al. (2014) define five core components that characterize a CURE. The following descriptions are adapted from their framework \cite{auchincloss_assessment_2014}.
\begin{itemize}
    \itemsep0em 
    \item \textbf{Use of scientific practices.} Students engage with multiple scientific practices, such as asking questions, building and evaluating models, proposing hypotheses, using the tools of science, gathering and analyzing data, developing and critiquing interpretations and arguments, and communicating findings. 
    \item \textbf{Discovery.} Students address novel scientific issues, developing and testing new hypotheses, and the students' results should offer new insight to the issue.
    \item \textbf{Broadly relevant or important work.} Students contribute to current scientific knowledge and have the opportunity for impact beyond the classroom.
    \item \textbf{Collaboration.} Students work collaboratively, practicing communication skills and integrating multiple perspectives into problem solving and addressing a shared research question.
    \item \textbf{Iteration.} Students engage in iterative processes, including revising experimental designs, refining and repeating analyses, and troubleshooting or problem solving in response to results.
\end{itemize}
These components do appear in other lab-course modalities, but all five components are required to distinguish a CURE from other course formats. While the CURE components are often discussed as distinct components, students may not experience these components independently in practice. From a sociocultural learning perspective \cite{vygotsky1978mind}, learning environments are shaped through interconnected experiences involving participation, collaboration, identity development, and engagement with disciplinary practices (e.g.~cite \cite{lave1991situated, van2011workplace, esteban2014funds,irving_sayer2014}). As a result, different aspects of authentic research experiences may reinforce one another in students’ reflections on their work.

Although physics-specific studies of CUREs remain limited \cite{dolan2016,buchanan2022}, a substantial body of research in other STEM disciplines, particularly biology, has examined their impact. Students participating in CUREs demonstrate increased self-efficacy and motivation and are more likely to persist in their fields of study \cite{gentile_undergraduate_2017}. Studies also report gains in content knowledge and analytical skills \cite{auchincloss_assessment_2014}, with outcomes comparable to those associated with traditional undergraduate research experiences (UREs) \cite{dolan2016}. Unlike traditional research experiences, which are typically available to only a small number of students, CUREs provide entire classes with opportunities to engage in research questions relevant beyond the classroom, while lowering barriers to participation \cite{bangera_course-based_2014}. Despite this growing body of evidence showing that CUREs promote positive student outcomes in other STEM disciplines, and the explicit recommendation of their implementation in the Effective Practices for Physics Programs (EP3) guide as a strategy for expanding access to authentic research experiences \cite{ep3}, physics has been identified as a field where CUREs remain limited within the curriculum \cite{buchanan2022}.

In the absence of established frameworks to support the development and sustainability of physics-specific CUREs, and given the amount of effort required to develop  them, instructors may be hesitant to create such courses. Additionally, due to the limited study of CUREs in physics, it is not clear if there are fundamental differences between the fields of physics and chemistry/biology that lead to additional challenges in creating and sustaining CUREs in physics, such as types of research questions or equipment. As a result, it remains unclear how best to expand CURE opportunities within physics or how to design them for long-term sustainability. Rigorous education research is needed to understand these discipline-based differences and build frameworks for implementing physics CUREs in a sustainable manner. Meeting this need has motivated our broader research efforts. 

The work presented here is part of our broader effort to support the implementation and sustainability of physics CUREs. Our long-term goals are to (1) identify the opportunities and challenges for implementing and sustaining physics CUREs and (2) develop a framework of effective practices to support the creation and expansion of CUREs in physics. This framework will address topics such as identifying strong research partnerships, designing for engagement with all five CURE components, assessing course impact, and preparing instructional teams for research-based instruction. As an initial step toward these broader goals, we developed, implemented, and evaluated a photovoltaics-focused CURE in a second-year undergraduate physics laboratory course at the University of Colorado Boulder. The CURE began in Fall 2024, with students investigating how environmental stressors (temperature and illumination) affect the external quantum efficiency of metal-halide perovskite (MHP) solar cells in partnership with the National Lab of the Rockies (NLR), formerly National Renewable Energy Laboratory.

Here, we make two complementary contributions toward these broader goals. First, we document the development, structure, and initial implementation of a scalable, in-person experimental physics CURE to provide a discipline-specific model that can support instructors interested in developing similar courses. Second, we analyze data from an end-of-semester reflection assignment to characterize student experiences during the first implementation of this CURE. Specifically, we investigate the following research questions (RQs):
\begin{enumerate}
    \item RQ1: How did students perceive and characterize their experiences with the five CURE components integrated into the course?
    \item RQ2: How did students describe and interpret their authentic research experiences, including perceptions of authenticity and engagement with ill-defined problems?
    \item RQ3: What immediate reflections did students report about their future academic and professional goals at the conclusion of the CURE?
\end{enumerate}
In the sections that follow, we first situate this work within the broader literature on CUREs. We then describe the design and implementation of the photovoltaics CURE, followed by an analysis of first-semester outcomes and implications for future iterations.

\section{Background}\label{Back}
To determine how physics CUREs can be implemented in scalable and sustainable ways, it is necessary to consider both the broader landscape of undergraduate research and the structural challenges that shape CURE adoption across STEM disciplines. In this section, we examine student benefits and access issues of traditional UREs, outline the documented benefits of CUREs and identified implementation challenges, and then examine the current state of CUREs in physics within this larger context.

\subsection{Benefits of, and barriers to participation in, UREs}
 Prior research has shown many benefits to student participation in traditional UREs,\cite{lopatto_survey_2004, lopatto_undergraduate_2007, thiry_what_2011, auchincloss_assessment_2014, gentile_undergraduate_2017} which we define as a student working with a faculty member, postdoc, or graduate student in an apprenticeship-style model.  UREs have been shown to improve student performance in STEM courses \cite{gentile_undergraduate_2017} and increase persistence in STEM degree programs \cite{jones2010}. Through research experiences, students are able to develop transferable research skills \cite{seymour_establishing_2004} and learn to think and work `like a scientist' \cite{hunter_becoming_2007, mrazcraig2018}. These opportunities allow students to begin developing their professional network by facilitating relationships with their peers and with senior scientists \cite{auchincloss_assessment_2014}. Students who participate in UREs have also demonstrated an increased interest in, and preparedness for, entering the STEM workforce\cite{trott_exploring_2020}. These outcomes have also been shown to be particularly beneficial to students from marginalized groups \cite{carlone2007, jones2010, thiry_what_2011, eagan2013, bangera_course-based_2014, pierszalowski_systematic_2021}. 

However, there are barriers that prevent many students from participating in traditional UREs. The demand for UREs often exceeds the number of available opportunities, and many departments lack the resources to support broad student participation  \cite{auchincloss_assessment_2014, hanshawperc2015}. Due to the limited number of research opportunities, strict selection criteria, such as minimum course grade requirements \cite{bangera_course-based_2014}, are commonly used to select undergraduate researchers. These criteria may not reflect a student's research capabilities and can result in capable students being excluded from research. Additional barriers include a lack of awareness about how to pursue research opportunities \cite{russell2007}, as well as other financial or  personal barriers (e.g., family responsibilities) that make participation in traditional UREs inaccessible \cite{pierszalowski_overcoming_2018,pierszalowski_systematic_2021}.

CUREs offer a potential way to lower these barriers and create opportunities for a larger population of students to engage in authentic research \cite{bangera_course-based_2014, rodenbusch2016early, dolan2016}. It should be noted that CUREs are not intended to replace traditional UREs, but rather to complement them. 

\subsection{Broad benefits of CUREs}
Participation in CUREs has been shown to result in similar student outcomes to those associated with UREs. Students demonstrate increased engagement with, and persistence in, STEM \cite{graham_persistence_2013, linn_undergraduate_2015, rodenbusch2016early}. They also show gains in scientific reasoning, critical thinking, and a deeper understanding of the nature of scientific research and the practices of working scientists \cite{auchincloss_assessment_2014, brownell2015high, rowland2016we}. CUREs help students build research and technical skills \cite{auchincloss_assessment_2014, corwin_effects_2018, werth_assessing_2022}, and foster increased science identity and project ownership \cite{cooper2019impact, hanauer2017}. Participation in CUREs has also been linked to greater science self-efficacy \cite{graham_persistence_2013, wilczek2022catalyzing, newell2022gains}, and can support students in clarifying and reinforcing their STEM career goals \cite{lopatto_undergraduate_2007, harrison2011classroom, corwin_effects_2018, newell2022gains}. Finally, CUREs provide opportunities for students to engage with more authentic measures of scientific success. Unlike traditional lab courses that often prioritize correct answers, CUREs emphasize iteration, problem-solving, and collaboration. Students report that successfully navigating obstacles, contributing to effective teamwork, and engaging deeply with research processes are meaningful outcomes of their CURE experience \cite{gin2018students, werth_assessing_2022, corwin2022students}.

\subsection{Challenges to designing and implementing CUREs in non-physics disciplines}
The positive impacts of CUREs are numerous, but it is also important to acknowledge the challenges associated with their development and implementation. Because CUREs remain underrepresented in physics, much of the existing literature on implementation barriers originates from biology education research. Brownell and Tanner, 2012 \cite{brownell2012barriers} identified three main barriers to implementing CUREs: lack of time, lack of incentives, and lack of training.

Instructors often face competing demands from research, teaching, service, and other professional responsibilities, making the time investment required for developing and implementing CUREs difficult \cite{brownell2012barriers, spell2014redefining, shortlidge2016faculty, govindan2020fear, orton2025challenges}. These time demands begin with learning a new teaching approach and continue through the course development and implementation phases. Even after a CURE is established, it typically requires more instructional effort than traditional lectures or prescriptive lab formats \cite{shortlidge2016assess}. Graduate TAs have also expressed concerns about the time commitment involved in learning and teaching as part of a CUREs \cite{heim2019benefits}. Beyond personal time commitments, instructors also express concern about aligning CURE content with existing course structures and timelines. \cite{lopatto2014central}.

These time constraints are exacerbated by the lack of incentives for instructors to pursue implementing CUREs. Instructors identified incentives that would encourage them to create and teach a CURE, including temporary reduction in their teaching load, financial support, and recognition of teaching efforts through awards and consideration for tenure and promotion \cite{brownell2012barriers}. However, the extent to which these incentives can be offered often depends on the institutional context. It appears to be rare for instructors to be explicitly prevented from developing a CURE, but support at both the departmental and administrative levels may still be limited \cite{govindan2020fear}.

Instructors and TAs also reported a lack of training as a barrier, both in terms of science content and pedagogical preparation \cite{brownell2012barriers, heim2019benefits}. When a CURE's research topic is outside an instructor's area of expertise, they may feel unprepared to support students and to guide graduate TAs effectively. For TAs, limited familiarity with the research content combined with minimal teaching experience can compound feelings of uncertainty and add to the challenge of facilitating a CURE. This, in turn, contributes to instructors' concerns about finding TAs who are adequately prepared to teach in these courses \cite{lopatto2014central}.

Beyond these three core barriers, instructors also report logistical and structural challenges. Instructors cite resource limitations, such as lack of funding for equipment or materials needed to complete the research, a lack of available space to conduct the research, and a lack of personnel, either graduate TAs or other instructors, as a challenge to implementing CUREs \cite{shortlidge2016faculty, govindan2020fear, orton2025challenges}. Limited resources may also contribute to faculty having difficulties finding an authentic research project that they can implement with what is available to them. Instructors may also struggle to find a project that aligns with the goals and constraints of the specific course being transformed.\cite{orton2025challenges}.

Finally, instructors and TAs also report student-related barriers to implementing CUREs. These include concerns about students' academic readiness and their willingness to engage with the structure and goals of a CURE. Because students are often accustomed to laboratory activities with predetermined outcomes, they may struggle with the inherent open-endedness and uncertainty of authentic research \cite{govindan2020fear}. This mismatch between student expectations and the nature of a CURE can lead to frustration or resistance. Instructors also express concern about students' ability to collaborate effectively \cite{shortlidge2016faculty}. While collaboration is central practice of scientific research, many students report disliking group work, despite typically reporting positive experiences after the fact \cite{chang2018group}, which can pose additional challenges in CURE settings.

\subsection{CUREs in physics}
\subsubsection{Representation of Physics CUREs in the Literature}
A comprehensive overview by Buchanan \& Fisher \cite{buchanan2022} identified 242 CUREs implemented across STEM disciplines between 2000 and 2020, with physics identified as one of the fields with relatively few published descriptions of CURE implementation. However, minimal does not mean non-existent. On CUREnet, two examples are listed: one focused on catastrophic cancellation in a traditional elastic collision experiment \cite{Walkup2020, walkcupcurenet}, and another in geophysics comparing methods for detecting cave features and mapping minor cave passages in the Bracken Preserve \cite{karstcurenet}. Additional CURE and ``CURE-like" programs in physics and astrophysics have explored topics such as imaging and reducing uncertainties in asteroid orbits \cite{wooten2018astrocure}, constraining orbital parameters of large transiting exoplanets \cite{hewitt2023}, building and testing instruments to study properties of a local river \cite{jensen2023}, and constructing muon detectors to investigate shielding effectiveness and variations in muon production due to solar activity and local weather conditions \cite{rabosky2025muoncure}. In response to the COVID-19 pandemic, CU-Boulder also developed and implemented the first remote, large-enrollment (N$>$400 per semester) physics CURE \cite{werth_impacts_2022, werth_assessing_2022, werth_eclass_2023, oliver2023}.

These examples represent a heterogeneous collection of instructional models. Existing examples include CURE-like or quasi-CURE astronomy experiences, authentic learning experiences not explicitly framed as CUREs, geophysics and astrophysics CURE implementations, and multiple publications arising from the same remote, large-enrollment solar physics CURE. Collectively, these studies demonstrate that physics CUREs can be implemented across a variety of contexts, but comparatively little has been published about the implementation of, and outcomes from, in-person experimental physics CUREs. Consequently, important questions remain regarding how these courses can be designed, sustained, and adopted more broadly.

\subsubsection{Barriers to Implementing CUREs in Physics}
To better understand how these challenges manifest in physics departments, we conducted interviews in Fall 2023 with 33 physics and astrophysics instructors across 19 institutions \cite{merritt2024}. Our findings revealed substantial overlap with the challenges reported in biology. The instructors interviewed identified time as the biggest challenge. They identified similar incentives, course releases and summer salaries, that would allow them to develop a CURE. Many expressed concerns about their preparedness to manage and support personnel, specifically graduate TAs. 

Additionally, physics instructors discussed concerns about students' `academic maturity.' Because students are often accustomed to laboratory activities with known outcomes, instructors worried they might struggle to adapt to the uncertainty and `messiness' of authentic experimental research. One instructor noted a concern that students might internalize their frustration at not arriving at a `correct answer' as a personal failure. A small number of instructors mentioned that limited physical resources, such as equipment and lab space, could present challenges. However, most instructors did not anticipate institutional resistance to implementing a CURE and believed that aligning the course with career readiness goals would make it broadly acceptable. At the same time, they expressed uncertainty about what concrete institutional support, such as funding, equipment, laboratory space, or teaching releases, would be available to sustain implementation\cite{merritt2024}.

These findings underscore the importance of institutional support, intentional course design, and personnel preparation in developing sustainable physics CUREs. Together, they highlight the structural considerations that must be addressed to support long-term implementation within physics departments.

\subsubsection{Colorado Physics Laboratory Academic Research Effort (C-PhLARE) CURE}
While these barriers present meaningful challenges to sustained implementation under typical institutional conditions, the COVID-19 pandemic created an opportunity to experiment with large-scale physics CURE design under remote laboratory constraints. In response, CU-Boulder developed the Colorado Physics Laboratory Academic Research Effort (C-PhLARE). The C-PhLARE CURE was a solar physics research experience that ran for three semesters (Fall 2020, Spring 2021, and Fall 2021). The project was mentored by a research scientist from the Laboratory for Atmospheric and Space Physics at CU-Boulder and aimed to investigate the heating mechanism of the solar corona. Two leading hypotheses, nanoflares and magnetohydrodynamic waves, have been proposed as sources of this heating \cite{oliver2023}. Students analyzed x-ray data from one of the Geostationary Operational Environmental Satellites (GOES-15) that was publicly available in the Space Weather Data Portal to examine if nanoflares could be frequent enough to account for the heating of the corona.\cite{knuth2020swx, lasp_space_weather_portal}    The more than 1200 students who participated in the C-PhLARE CURE contributed the results, which are detailed in Mason et al., 2023 \cite{mason2023}, with students as co-authors.

For this remote CURE, students were put into teams of 3--4 and met synchronously weekly with their team for two hours via Zoom. A graduate TA also attended these synchronous meetings. The C-PhLARE CURE consisted of the following phases \cite{werth_assessing_2022}:  (1) Project onboarding, (2) Research plan development, (3) Data analysis, (4) Peer review, (5) Determining power-law of distribution, (6) Documentation and reflection. Each  team of students selected and analyzed at least one solar flare from the Space Weather Data Portal \cite{knuth2020swx, lasp_space_weather_portal}. The results from the students' analysis were then aggregated to create the flare frequency distribution. Additional details outlining the course phases can be found in Werth et al., 2022 \cite{werth_assessing_2022}.

\subsubsection{Success of C-PhLARE CURE}
In addition to contributing meaningfully to the solar physics community, the C-PhLARE CURE positively impacted participating students. Many students reported engaging in authentic research practices, including using scientific tools, answering research questions, and sharing their results with the scientific community \cite{oliver2023}. Students also reported gains in coding skills and confidence using coding as a research tool \cite{werth_impacts_2022}. Positive teamwork was another significant impact of the course. Students described their group experiences as both productive and enjoyable, and identified collaboration as an essential part of their overall research experience \cite{werth_assessing_2022}. Analysis of the Colorado Learning Attitudes about Science Survey for Experimental Physics (E-CLASS) \cite{eclass} responses further showed that C-PhLARE participation increased students' expert-like views related to authentic research, science communication, and confidence in their research abilities \cite{werth_eclass_2023}.

These outcomes highlight the impact of a physics CURE. However, while the C-PhLARE project successfully documented course-level impacts and offered guidance for implementing remote CUREs, it did not, and was not meant to, address long-term sustainability of CUREs \cite{werth_impacts_2022}. Because it was designed to operate only while remote labs were required during the pandemic, questions remain regarding how to design physics CUREs that are sustainable, scalable, and compatible with in-person laboratory environments. Furthermore, due to its remote structure, the C-PhLARE CURE could not investigate the opportunities and challenges associated with in-person lab settings. To help fill these gaps, we have developed an in-person physics CURE in which students investigate how environmental stressors affect the performance of MHP photovoltaics.

\section{Course Development}

The transformed course, PHYS~2150: \emph{Experimental Physics 2}, is a one-credit, second-year lab course at CU-Boulder, which was previously offered as a traditional prescriptive modern physics laboratory course. The course format consists of a weekly 50-minute lecture and a two-hour lab meeting. The lecture component ends after the first half of the semester. PHYS~2150 is a required course for physics, astrophysics, and engineering physics majors and serves $\sim$200 students per academic year. While it is not directly linked to any theory lecture course, PHYS~2130: \emph{Introduction to Quantum Mechanics and Its Applications} or PHYS~2170: \emph{Foundations of Modern Physics} are prerequisite or corequisite courses for students taking this lab. Additionally, PHYS~1140: Experimental Physics 2, which is an introductory lab course, is a prerequisite.

The development of the PHYS~2150 CURE began with instructor interviews. We conducted interviews with 33 instructors, representing a variety of physics research specializations, from 19 different institutions, ranging from community colleges to R1 institutions. Eleven of the instructors were from CU-Boulder. The instructors were a combination of research faculty, teaching faculty, and lab coordinators. The purpose of these interviews was to probe instructor views on their learning goals for lab courses and what they identified as challenges and opportunities of implementing CUREs \cite{merritt2024}. From these interviews, error analysis/propagation, data analysis, experimental design, programming, and writing were the most commonly identified learning goals \cite{merritt2024}. We also hosted an open meeting with CU-Boulder physics faculty to discuss the PHYS~2150 course transformation and gather additional input on learning goals. The faculty identified measurement uncertainty/error analysis and hands-on equipment use/data acquisition as the top two priorities. 

Based on the discussions with faculty, the context of PHYS~2150 in the broader department curriculum, and the CURE framework, the following learning goals were chosen:

Students should be able to...
\begin{itemize}
    \itemsep0em 
    \item explain the main components of the research process
    \item design and execute an experimental plan
    \item use Python to do basic experimental data analysis
    \item identify a research question and conduct analysis to answer it
    \item iterate on proposals and experimental design based on feedback
    \item work collaboratively with a team of students to achieve a common goal
\end{itemize}
It was equally important to make clear to the students and instructors what were \emph{not} learning goals for the course. In the syllabus, students are told that this course is not meant to:
\begin{itemize}
    \itemsep0em     
    \item reinforce physics concepts covered in the pre- or corequisite courses.
    \item make you an expert python programmer.
    \item have you understand all of the details of solar cells.
\end{itemize}
Additionally, the course includes metacognitive goals, alongside personal and professional development goals. These include clarifying career aspirations, strengthening students’ identities as scientists and members of the scientific community, and helping students recognize their participation in authentic research, particularly by engaging with the uncertainty, iteration, and complexity inherent to experimental science.

Once the goals for the course were established, a research focus and associated science mentor had to be identified. Sustainability was a central consideration in selecting the research component of the course, and several constraints guided both decisions. In this context, sustainability referred not only to financial feasibility, but also to logistical, instructional, and research continuity across semesters.

The project needed to be scientifically relevant, such that students would contribute new insight to the research area, while remaining feasible within the time structure of a semester-long laboratory course. Additionally, the project design also needed to make clear why student participation was necessary, ensuring that students were contributing meaningfully to the research rather than completing tasks that could be more efficiently performed by a different team of researchers or automated analysis. The course needed to rely on equipment and materials accessible within existing departmental resources and financial constraints, be scalable to accommodate all enrolled students, and generate datasets that were meaningful, yet manageable, for novice researchers. The project therefore had to support repeated implementation across semesters without requiring substantial modification to equipment or specialized expertise beyond the instructional team.

In addition to these considerations, the science mentor needed to be willing to engage not only with the instructional team, but also directly with students across multiple semesters. We define the science mentor as an active researcher within the research domain of the CURE who collaborates with the instructional team and students to provide scientific direction, ensures that the project reflects the norms and methodologies of that subfield, and supports the integration of the research project into the broader scientific community. This required a commitment to participating in course activities as needed, including visiting the class, assisting with background literature and scientific framing for both instructors and students, and providing essential research materials or tools, such as experimental samples or commonly used data analysis tools and methods.

Within our institutional context, the partnership with NLR on studying photovoltaic degradation met these considerations in several key ways. In addition to being a senior research scientist at NRL, the science mentor is also physics faculty at CU-Boulder. This made him accessible to both the instructional team and the students, while the research topic itself satisfied many of the scientific and logistical requirements of the course. MHP devices represent an active and rapidly evolving area of materials research (e.g., \cite{li2018scalable, huperovskite2023, olasoji2025metal, shen2026key}), providing opportunities for authentic student contributions. The range of parameters to test (i.e., wavelength, temperature, fabrication method) was too large to be explored by NRL scientists in a reasonable amount of time; thus, the contribution from a large number of students was critical and not artificial. The experimental measurements required to probe device performance could be completed within a single semester and were readily scalable across multiple student teams. The equipment needed for the experiments was not overly expensive or complicated and could be easily fabricated and/or purchased.  Additionally, the science mentor’s ongoing research program and sustained engagement with the course ensured continued access to materials and a sustainable overarching research direction, with flexibility to pursue distinct semester-specific research questions  in future iterations within the same course structure and with the same equipment. This partnership further enabled the course to focus on a research problem with clear scientific and societal relevance, which included renewable energy through the use of photovoltaics.

Photovoltaics are an important component of the transition to renewable energy. MHP devices, first developed for photovoltaics applications only about a decade ago, have emerged as promising materials because they can achieve power conversion efficiencies comparable to silicon while offering significantly lower energy payback times due to simpler fabrication processes. However, MHP devices currently suffer from lower stability and reliability than conventional silicon photovoltaics. To better understand the factors that influence device degradation, this project investigates how performance changes as a function of illumination wavelength and temperature for devices fabricated using different material compositions and fabrication methods. Because many of the degradation pathways in these materials remain poorly understood, empirical measurements are essential for identifying the mechanisms that limit device lifetime. To quantify the device performance, we are using the industry standard measurements, current density vs. voltage (J-V) and external quantum efficiency (EQE). 

\begin{figure}[h!]
    \centering
    \includegraphics[width=\linewidth]{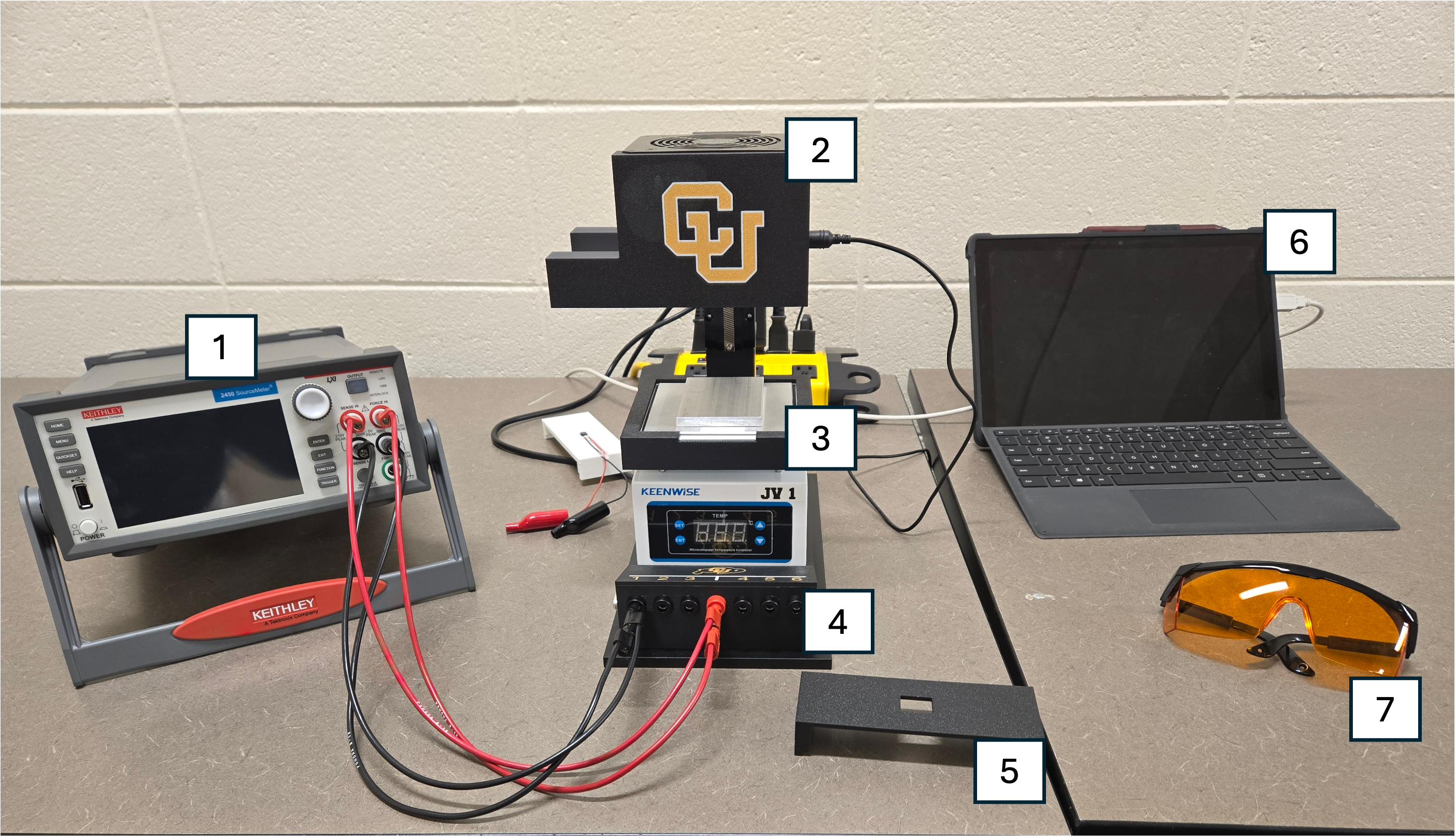}
    \caption{Current density-Voltage (J-V) apparatus. (1) Sourcemeter, (2) Solar Simulator, (3) Hotplate, (4) Electrical connection breakout box, (5) Alignment device, (6) Computer with user interface, (7) UV protection glasses}
    \label{fig:jv_apparatus}
\end{figure}

\begin{figure}[h!]
  \centering
  \includegraphics[width=.9\linewidth]{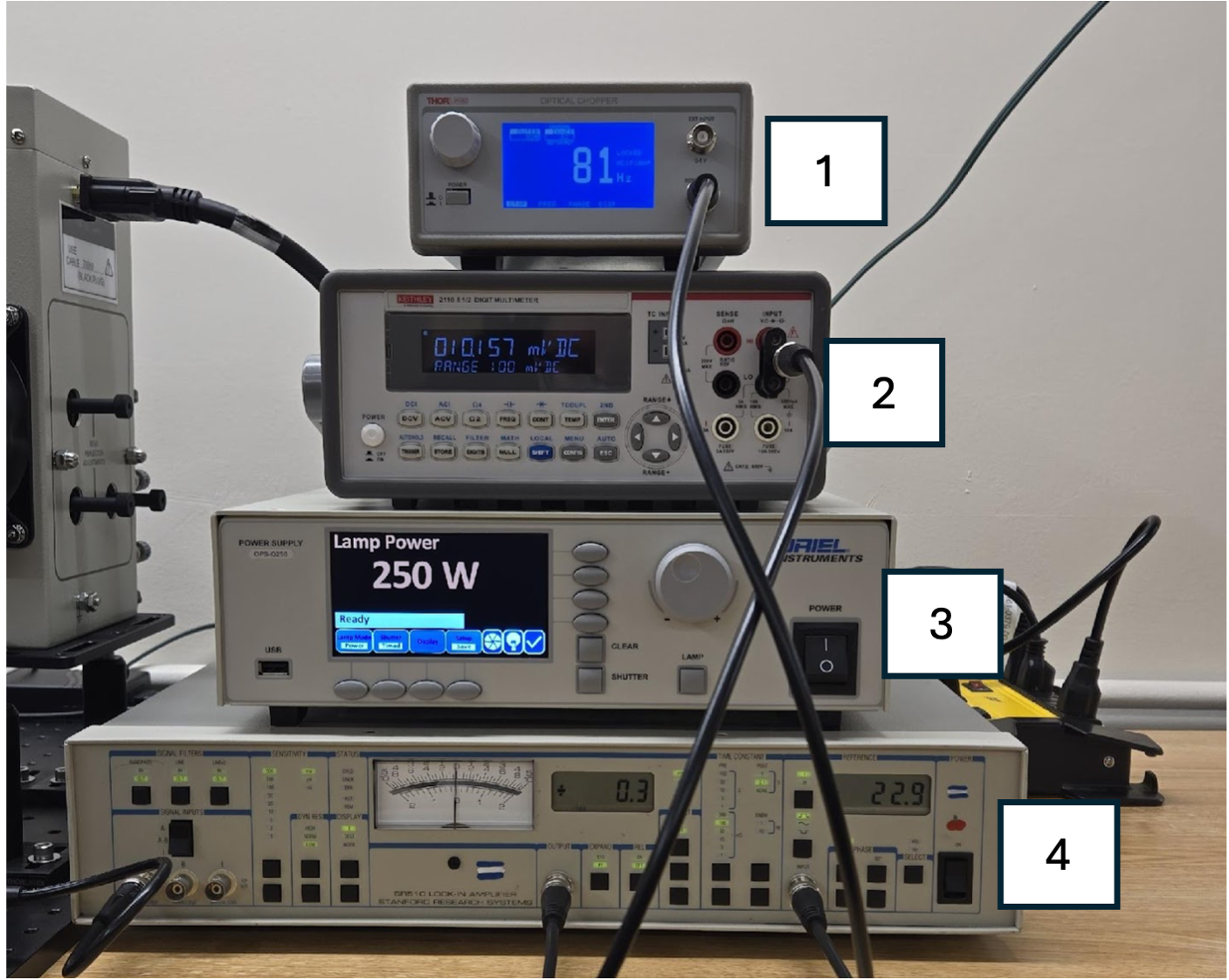}
  \vspace{0.5em}
  \includegraphics[width=.9\linewidth]{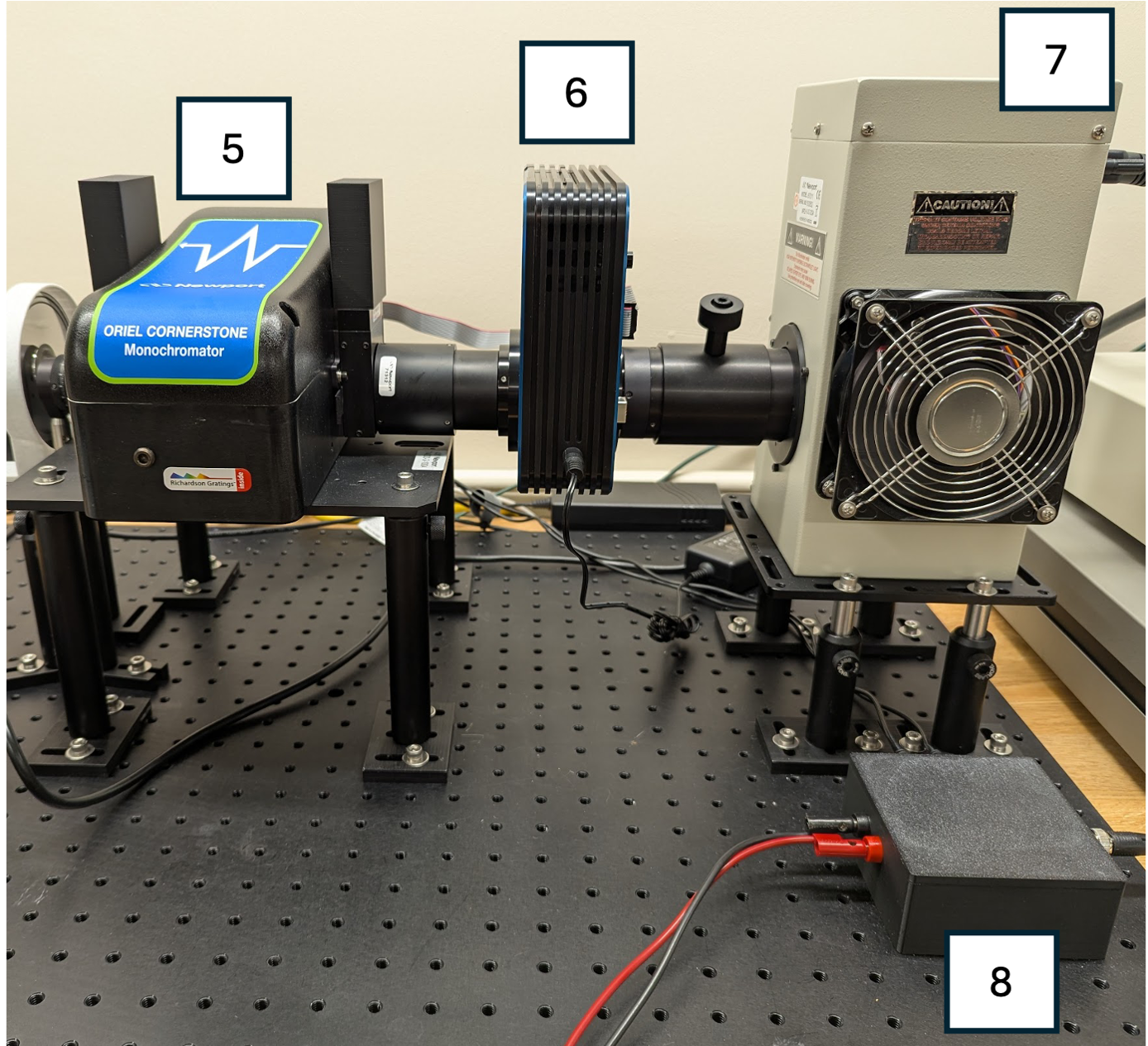}
  \vspace{0.5em}
  \includegraphics[width=.9\linewidth]{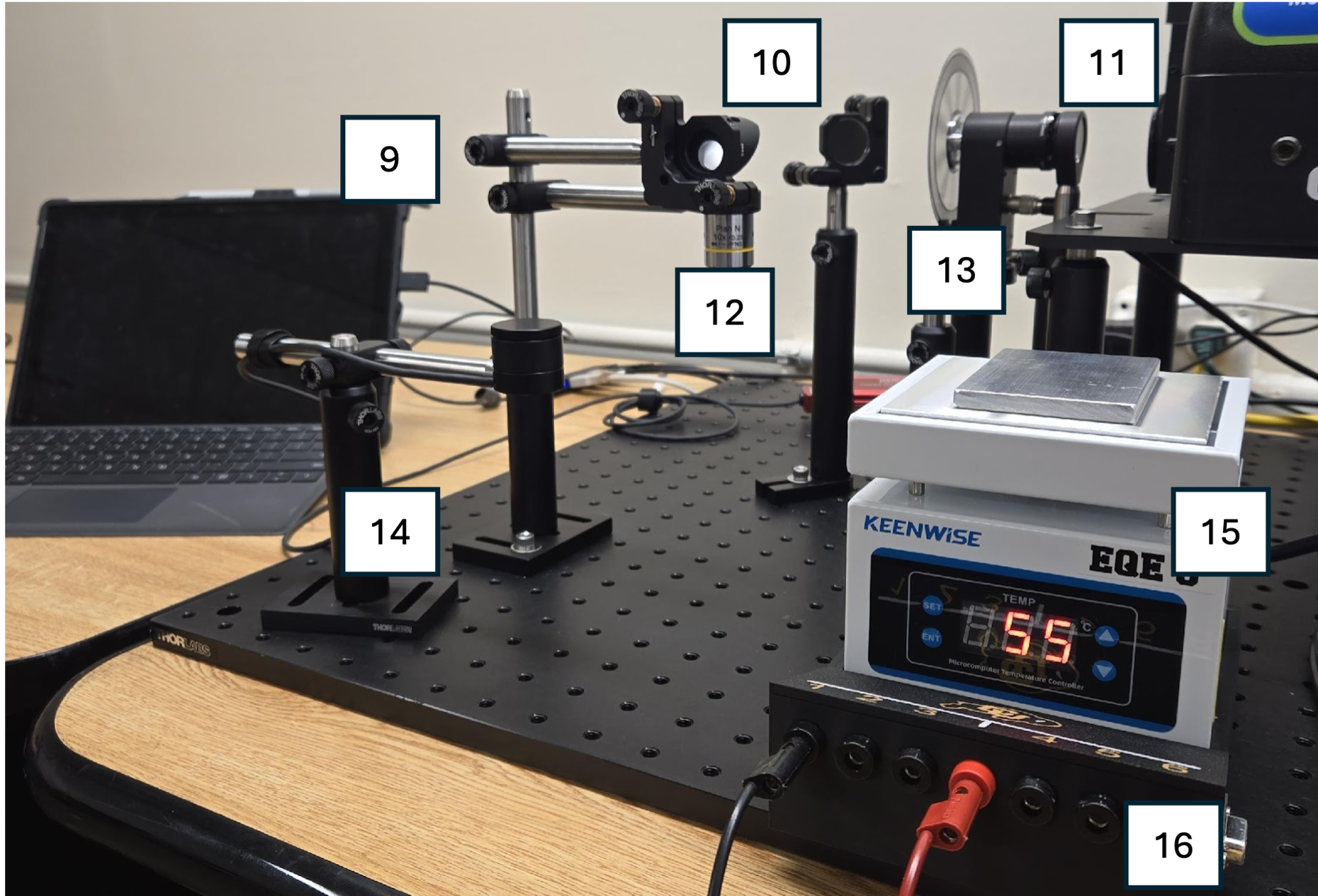}
  \caption{External quantum efficiency (EQE) apparatus. (1) Chopper-wheel controller, (2) Digital multimeter, (3) Lamp power supply, (4) Lock-in amplifier, (5) Monochromator, (6) Filter wheel, (7) Quartz tungsten halogen lamp, (8) Transimpedence amplifier, (9) Computer with user interface, (10) Mirrors, (11) Iris, (12) Microscope objective, (13) Chopper wheel, (14) Power-meter sensor, (15) Hotplate, (16) Electrical connection breakout box}
  \label{fig:eqe_apparatus}
\end{figure}

Two data collection apparatuses, one for measuring (J-V) and another for measuring (EQE), were built by our Director of Undergraduate Laboratories the summer before the course ran for the first time. These apparatus were modified and improved over the following year based on feedback from the students and graduate TAs.  Labeled images of the apparatus can be seen in Figures~\ref{fig:jv_apparatus} and \ref{fig:eqe_apparatus}.

\section{Course Structure}
\begin{figure*}[h!]
    \centering
    \includegraphics[width=\linewidth]{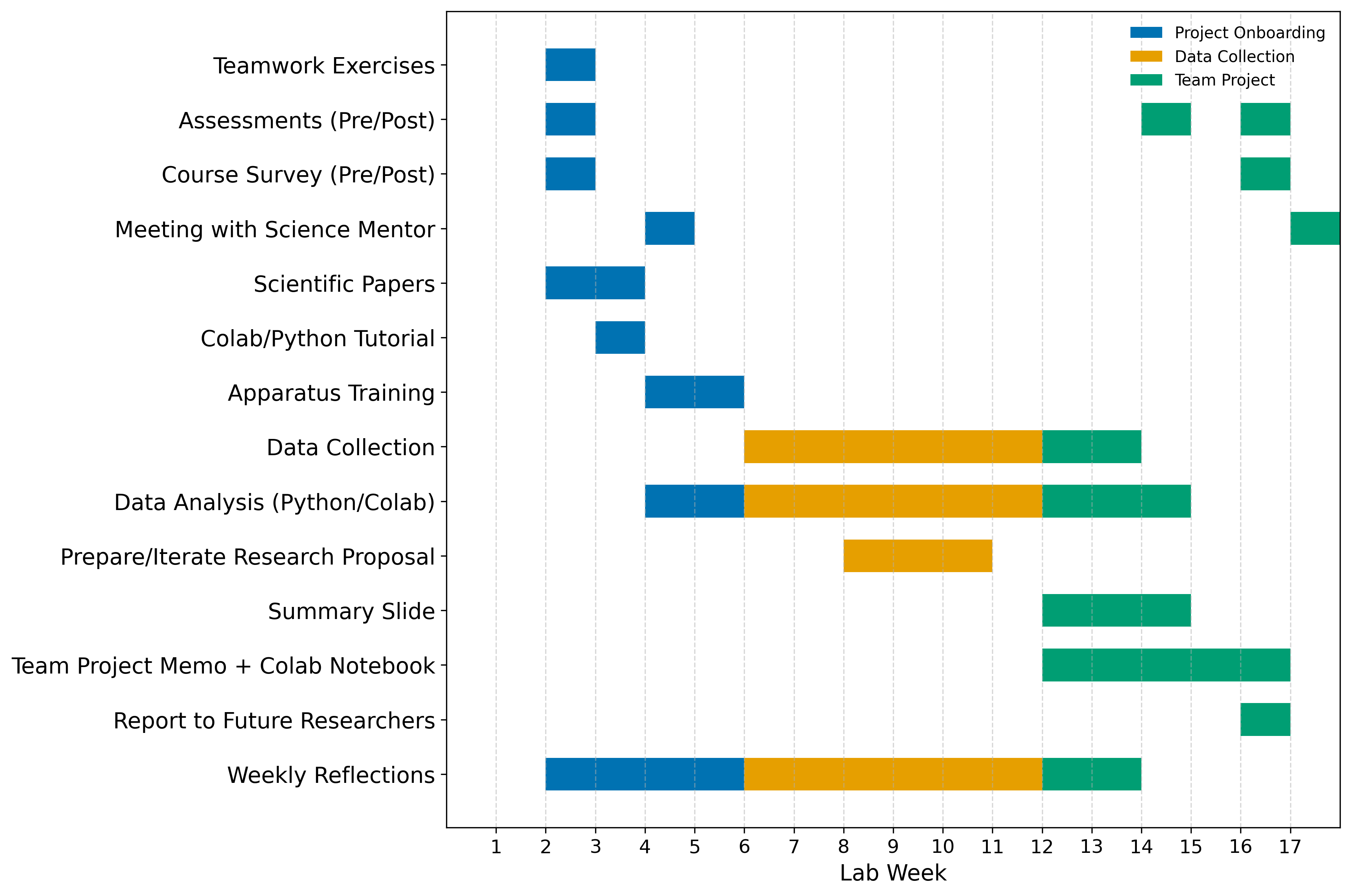}
    \caption{Timeline of course activities. The three phases of the course are distinguished by color: Research onboarding (Weeks 1–5) -- blue, data collection (Weeks 6–11) -- orange, and investigating team project questions (Weeks 12–17) -- green. This timeline reflects the Fall 2025 semester following the introduction of a university reading day. These activities were also implemented during the Fall 2024 semester, though the timing of activities differed slightly between semesters.}
    \label{timelinefig}
\end{figure*}

The course structure  consists of three phases, (1) Research Onboarding, (2) Data Collection and analysis, and (3) Investigating Team Project Question. Each component takes approximately one-third of the semester. A timeline of the course is shown in Fig.~\ref{timelinefig}. Detailed descriptions of course activities and the course schedule, including deliverables, are provided in the Supplemental Materials \cite{suppmat}.

The course is structured around two complementary levels of research questions. At the course level, data collection is guided by a central research question defined in collaboration with the science mentor, ensuring that all student teams contribute to a shared, scientifically relevant dataset. At the team level, students develop their own research questions later in the semester using this aggregated dataset. This structure is intentional. Because students are new to research, they are not expected to formulate well-posed research questions prior to data collection. Instead, the course emphasizes the practice of generating new questions from existing data, which is a common approach in many areas of physics and astronomy (e.g., data mining and archival analysis).

\subsection{Lecture, Lab, and Assessment}
Across all three phases of the course, students participate in both lecture and laboratory components. During lecture time, there are seven interactive sessions on topics including the science of photovoltaics, measurement techniques and instrumentation, developing research questions and proposals, measurement uncertainty, fitting, collaboration, and data management. Lecture time also includes two group meetings with the science mentor (one at the beginning and one at the end of the semester), as well as a proposal workshop in which students work in teams to define their own research questions and analysis plans for the team project. 

During weekly lab session, students are provided with a lab guide that outlines the course activities for the week, including a suggested schedule and the deliverables due at the end of the week. As students become more familiar with the course and gain confidence in their work, the guides become progressively less detailed.

Coursework points are split approximately 50/50 between individual and team assignments (see Table~\ref{tab:assignments} for a detailed breakdown). While students are encouraged to discuss ideas and questions about individual assignments with their teammates, TAs, or the instructor, the work they submit must be their own. Team assignments are completed collaboratively, and each team member is required to submit an identical copy. The weekly Google Colaboratory (Colab) notebooks include an author contributions section, where students list each team member and briefly describe their contributions during that week's lab session. Example contributions include taking measurements, performing data analysis, and answering questions based on the week’s results. The use of artificial intelligence (AI) tools is permitted and encouraged for coding tasks, such as generating code snippets, debugging, and seeking coding-related advice. However, any AI-generated code must be reviewed by the student. Each Colab notebook includes an AI statement, where students are required to identify the AI-assisted tools they used, how they used them, and at what stage(s) of the coding process (e.g., initial development, troubleshooting) the tools were applied. Across the entire semester, students complete weekly reflections consisting of a mix of Likert scale and open-response questions. These reflections are meant to capture students' experiences and engagement with the five CURE components, as well as their metacognitive, personal, and professional development. 

\begin{table}[!t]
\centering
\caption{Overview of individual and team assignments and their associated point values in PHYS~2150. Additional details can be found in Supplemental Materials \cite{suppmat}.}
\label{tab:assignments}

\begin{tabular}{lccc}
\textbf{Assignment} & \textbf{Items} & \textbf{Points} & \textbf{Total pts} \\
\hline

\multicolumn{4}{c}{\textit{Individual Assignments (215 pts)}} \\
Lab Participation & 13 & 5 & 65 \\
Lecture Clicker Questions & 5 & 10 & 50 \\
Online Assessments / Surveys & 5 & 2 & 10 \\
Reflection Questions & 12 & 5 & 60 \\
Python Tutorial & 1 & 15 & 15 \\
Apparatus Video & 1 & 5 & 5 \\
Report to Future Researchers & 1 & 10 & 10 \\

\hline
\multicolumn{4}{c}{\textit{Team Assignments (175 pts)}} \\
Teamwork Scenarios & 1 & 5 & 5 \\
Colab Notebooks & 7 & 15 & 105 \\
Questions for Science Mentor & 1 & 5 & 5 \\
Data Management Plan & 1 & 5 & 5 \\
Team Project Proposal (2 iterations) & 2 & 10/5 & 15 \\
Team Project Results Slide & 1 & 10 & 10 \\
Team Project Colab Notebook & 1 & 5 & 5 \\
Team Project Results Memo & 1 & 25 & 25 \\

\end{tabular}%

\end{table}

\subsection{Research Onboarding}

The research onboarding phase is meant to prepare students for subsequent course activities by emphasizing teamwork, reading scientific literature, and hands-on experience with data collection and analysis. 

During the first lab meeting, the students work through a selection of teamwork scenarios inspired by the Center for the Improvement of Mentored Experiences in Research \cite{cimer}. The scenarios focus on communication and navigating through conflict. After working through their selected scenarios, each team suggests two norms that will encourage productive teamwork throughout the semester. In the first week, students also complete a pre-course survey and the pretests for two research-based assessment instruments: Colorado Learning Attitudes of Science Survey for Experimental Physics (E-CLASS) \cite{eclass} and the Survey of Physics Reasoning on Uncertainty Concepts in Experiments (SPRUCE) \cite{spruce}. 

During the second week of onboarding, students are given their semester-long team assignments. Teams are determined by the instructor and based on information collected in the pre-course survey, such as coding experience. We have found the ideal number of students per team is three, but this depends on the complexity of the research project. Teams of two students can be problematic if one student is not able to attend the lab session and also limits the students' opportunity to develop teamwork skills. 

Prior to the second lab session, students are assigned annotated sections of two scientific papers to read in preparation for the upcoming lab meeting with the science mentor. After being assigned to their teams, the students work together to summarize the papers and identify any unresolved questions. Each group shares their discussion summaries and questions with the full lab section, which has up to 18 students. After the section discussion, each team submits three questions for the science mentor. The questions are aggregated, curated, and sent to the science mentor ahead of their first lab visit. For the remainder of the second lab session, each student independently completes a tutorial designed to build foundational skills in Python and Colab. Based on the pre-course surveys, 24\% of students reported having zero coding experience prior to the course and 38\% reported feeling `not confident' with coding or programming. The tutorial, and subsequent notebooks, provide students with code templates that they are able to use throughout the semester (e.g., data structure, basic scripts for calculations and plotting, dataframes). 

The remaining two weeks of onboarding (Weeks 4 and 5) are spent on apparatus training and baseline (unstressed cells) data collection and analysis. Each team is given a device (cell) with 6-8 individual solar PV components (known as pixels) and they spend one week familiarizing themselves with each apparatus and taking data for each pixel. After collecting their data, students analyze it based on the measurement performed that week. For EQE measurements, they calculate and plot the external quantum efficiency curves. For current density–voltage measurements, they plot the J–V data and determine the power conversion efficiency of the device.

\subsection{Data Collection}  
\begin{figure}[h!]
  \centering

  \begin{minipage}[t]{0.48\textwidth}
    \centering
    \includegraphics[height=2.4in]{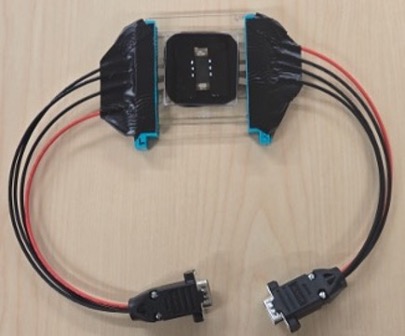}
  \end{minipage}
  \hfill
  \begin{minipage}[t]{0.48\textwidth}
    \centering
    \includegraphics[height=2.5in]{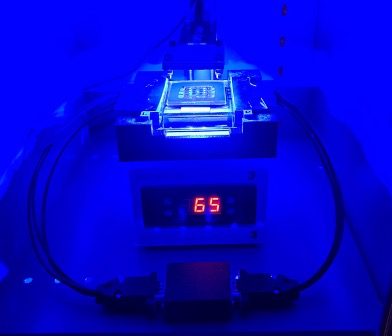}
  \end{minipage}

  \caption{Cell and stressing station. Top image: Example of solar cell with attached edgecards, which make electrical contacts with each individual pixel. Six white dots in the center of the cell packaging are guides to indicate location of pixels. Bottom image: Example of stressing station. A cell is located on a hotplate set to 65$^{\circ}$C with 450 nm wavelength LED illumination.}
  \label{fig:cellstress}
\end{figure}

The second phase of the course focuses on data collection and analysis while the cells are being continually stressed. During the Fall 2024 semester, students were provided with four temperature conditions (55, 60, 65, and 70 $^{\circ}$C) and four illumination wavelengths (450, 530, 595, and 730 nm), and each team selected the parameters under which their device would be stressed. A solar cell and stressing station are shown in Figure~\ref{fig:cellstress}.
Each week during the data collection phase, the students remove their team's cell from its stressing station and take both EQE and J-V measurements for each pixel. Students receive a Colab notebook each week to engage them with new analysis tools, including techniques for plotting from multiple CSV files, calculating and visualizing error bars, and performing linear regression and other fitting methods. This pattern continues for all weeks in the data collection phase.

After three weeks of data collection, students begin developing a team project question and analysis plan that they will complete using the aggregated class dataset. To support this process, students receive a structured proposal guide to help them organize their ideas. One lecture period is dedicated to a proposal workshop in which the instructor provides real-time feedback and answers questions as teams brainstorm their projects. Students then submit a first draft of their proposal, receive TA feedback, and revise and resubmit it for final instructor feedback.

\subsection{Investigating Team Project Question}
In the final phase of the course, students answer their team project question and communicate their results. The aggregated dataset from all teams is saved as an hierarchical data format version 5 (H5) file, and students spend three lab sessions completing their proposed analysis and working on presenting their results. Their results are communicated in two formats. First, each group creates a one-slide summary of their findings. These slides are compiled and shared with the science mentor prior to the final group meeting held on during the final lecture of the semester. This allows the science mentor to highlight the students' work and discuss how the data they collected over the semester will be used. Each team also writes a research memo that includes an overview of their research question, a description of their data selection and analysis methods, a presentation of their findings and interpretation, and a concluding section with key takeaways and suggestions for future work. The memo also includes author contributions and the Colab notebook used for the final analysis. In addition to the research memo and summary slide, students also complete the assessment posttests (E-CLASS and SPRUCE), a post-course survey, an end-of-course reflection assignment, and the \textit{Report to Future Researchers} assignment during the final weeks of the semester. 

\subsection{Report to Future Researchers}
In the final week of the semester, students completed a 400~$\textendash$~800 word \textit{Report to Future Researchers} (hereafter, reports). For this assignment, students were reminded that the research in the course would continue into the next semester and were asked to introduce the incoming students to the course and the project. Students were prompted to address the following questions:
\begin{enumerate}
    \item Explain the Research - How would you describe the purpose and significance of the project to someone new? How does each student's contribution fit into achieving the goals?
    \item Reflect on Lessons Learned - What were the most valuable lessons or skills you developed this semester, and how might they benefit future participants in the course?
    \item Advice for Future Participants - What strategies would you recommend to new students to help them effectively engage with the research process and contribute to the project?
    \item Keys to Success - What should a new student consider to succeed in both their individual work and the overall class project?
\end{enumerate}
The reports are incorporated into subsequent iterations of the course. During the first week of each semester, students are given access to all deidentified reports from the previous semester. Within their teams, each student randomly selects two reports to read individually, after which groups discuss the following prompts:
\begin{enumerate}
\item What themes do you see in these reports?
\item What surprises you in students' reflections on the course?
\end{enumerate}
This activity is designed to continue across semesters, with each cohort engaging with reports produced by the previous class. These data will be the focus of this analysis.

\section{Education Research Methodology}\label{methods}
\subsection{Study Participants}
The data used in this work were collected from students in the Fall 2024 semester of the CURE, which was the first time the course ran. During this semester, there were 127 students enrolled in the course. Demographic information was not collected as part of the reports; instead, demographic data are reported for 125 students based on pre-course survey responses (Table~\ref{tab:demographics}). Out of the students enrolled in the course, 108 submitted a report ($\sim$~85\%). Prior to analysis, the reports were deidentified by removing any names, lab section information, or other identifiable details from the text of the document. Additionally, the metadata fields in the description tab of the PDF files' document properties were cleared to remove any remaining identifying information.

\begin{table}[t]
\caption{Demographics of students enrolled in PHYS~2150 in Fall 2024 based on pre-course survey responses ($N=125$). Percentages are calculated relative to the number of respondents. Counts for \textit{Declared Major} do not sum to 125 because students were allowed to report multiple majors.}
\label{tab:demographics}
\setlength{\tabcolsep}{8pt}
\begin{tabular}{lcc}
\toprule
Category & Number of students & \% \\
\midrule

\multicolumn{3}{l}{\textit{Time at CU}} \\
First year & 6 & 4.8 \\
Second year & 75 & 60.0 \\
Third year & 30 & 24.0 \\
Fourth or beyond & 10 & 8.0 \\
Other & 4 & 3.2 \\

\midrule
\multicolumn{3}{l}{\textit{Gender}} \\
Man & 85 & 68.0 \\
Woman & 36 & 28.8 \\
Non-binary & 3 & 2.4 \\
Prefer not to answer & 1 & 0.8 \\

\midrule
\multicolumn{3}{l}{\textit{Declared Major}} \\
Physics & 53 & 42.4 \\
Astrophysics & 43 & 34.4 \\
Math/Applied Math & 26 & 20.8 \\
Engineering Physics & 21 & 16.8 \\
Engineering & 11 & 8.8 \\
Computer Science & 4 & 3.2 \\
Astronomy & 4 & 3.2 \\
Other science & 2 & 1.6 \\
Geology/Geophysics & 1 & 0.8 \\
Chemistry & 1 & 0.8 \\
Biology & 1 & 0.8 \\

\bottomrule
\end{tabular}
\end{table}

\subsection{Frameworks that informed analysis}
Because CUREs are intended to provide authentic research experiences, this work draws on prior work to examine different dimensions of students' experiences. The five CURE components described above provide the framework used to examine how students perceived and characterized the intentionally designed features of the course (RQ1). Prior work examining  how students perceive and experience authenticity with in CUREs provides a separate framework for examining student interpretations of their authentic research experiences (RQ2).

To distinguish CUREs from inquiry-based laboratory courses, Goodwin et al. (2021) identified several coded elements that contributed to students perceiving their research experiences as ``real" or authentic \cite{goodwin2021science}, which incudes students identifying as a real scientist or researcher. These elements included autonomy, failure, collaboration, iteration, scientific practices, relevant discovery, and successful science. Many of these elements overlap with established CURE components, particularly collaboration, iteration, scientific practices, and discovery. Building on these ideas, Oliver et al. (2023) created a coding scheme associated with perceptions of authenticity and applied it to student reflections in the remote C-PhLARE CURE, including direct authenticity, indirect authenticity, and failure \cite{oliver2023}. The initial codebook for this work was informed by both bodies of prior work while maintaining their distinct analytic purposes: the CURE framework informed analysis of students' experiences with intentionally designed course features for RQ1, while the authenticity framework informed analysis of students perceptions of authentic research for RQ2. 

\subsection{Data Analysis}
The reports were analyzed using a standard qualitative thematic coding process \cite{qualresearch}. Our initial codebook consisted of \textit{a priori} codes aligned with the five CURE components, as well as elements of authentic research as defined by Goodwin et al., (2021) \cite{goodwin2021science} and adopted by Oliver et al. (2023) \cite{oliver2023}. There were two major changes to the codebook in this work compared to the codebook used in Oliver et al. First, we did not include their `Indirect Authenticity' code because the subcodes it encompassed substantially overlapped with the CURE component of \emph{Relevance}. In Oliver et al., indirect authenticity captured students’ perceptions that their work contributed to broader scientific goals or audiences. In this work, those themes were instead categorized under \emph{Relevance}, which was coded as distinct from authenticity. We did, however, adopt their `Direct Authenticity' code under the revised name `Explicit Authenticity.' Second, rather than adopting the standalone `Failure' code, we have included it as a subcode within `Ill-defined Problems' to capture these direct references to failure.  We introduced an `Ill-defined Problems' code using the definition from Schraw et al.~\cite{schraw1995cognitive}: problems that have multiple and/or non-guaranteed solutions and lack a defined procedure for reaching them. This captured how students framed setbacks, uncertainty, and changing plans as part of the broader process of authentic research. Within this context, instances of experimental failure were understood as one possible outcome of engaging with ill-defined problems rather than as an isolated endpoint disconnected from the broader research process. In subsequent passes through the data, emergent subcodes were added under the main codes as needed. Definitions and example quotes for the main codes and subcodes used in this work are provided in the Supplemental Materials \cite{suppmat}.

Using eight reports, author RLM and a researcher in the larger research group not working on this project conducted an initial round of inter-rater reliability (IRR) coding using the codebook. Cohen’s kappa was calculated to be 0.91, which is near perfect agreement \cite{cohenkappa}. After this initial round, author RLM and the other researcher discussed and resolved the discrepancies. Because only minimal changes were made to the codebook and the pre-discussion kappa was already acceptable ($>$0.8), the IRR process was not repeated. Author RLM then coded the remaining reports. The unit of analysis in this study was one sentence. Because students often expressed multiple relevant ideas within a single sentence, individual sentences could be assigned multiple subcodes. While coding was conducted at the sentence level, results were aggregated and reported at the student level. Percentages reported for main codes represent the proportion of all 108 student reports assigned that code. Percentages reported for subcodes are calculated relative to the subset of reports coded under the corresponding main code. Because responses could receive multiple subcodes within a main code, subcode percentages do not necessarily sum to 100\%. With the exception of `Discovery' and `Career Clarification,' responses were coded at both the main code and subcode levels.

\section{Results}
\begin{figure*}[t]
    \centering
    \includegraphics[width=\textwidth]{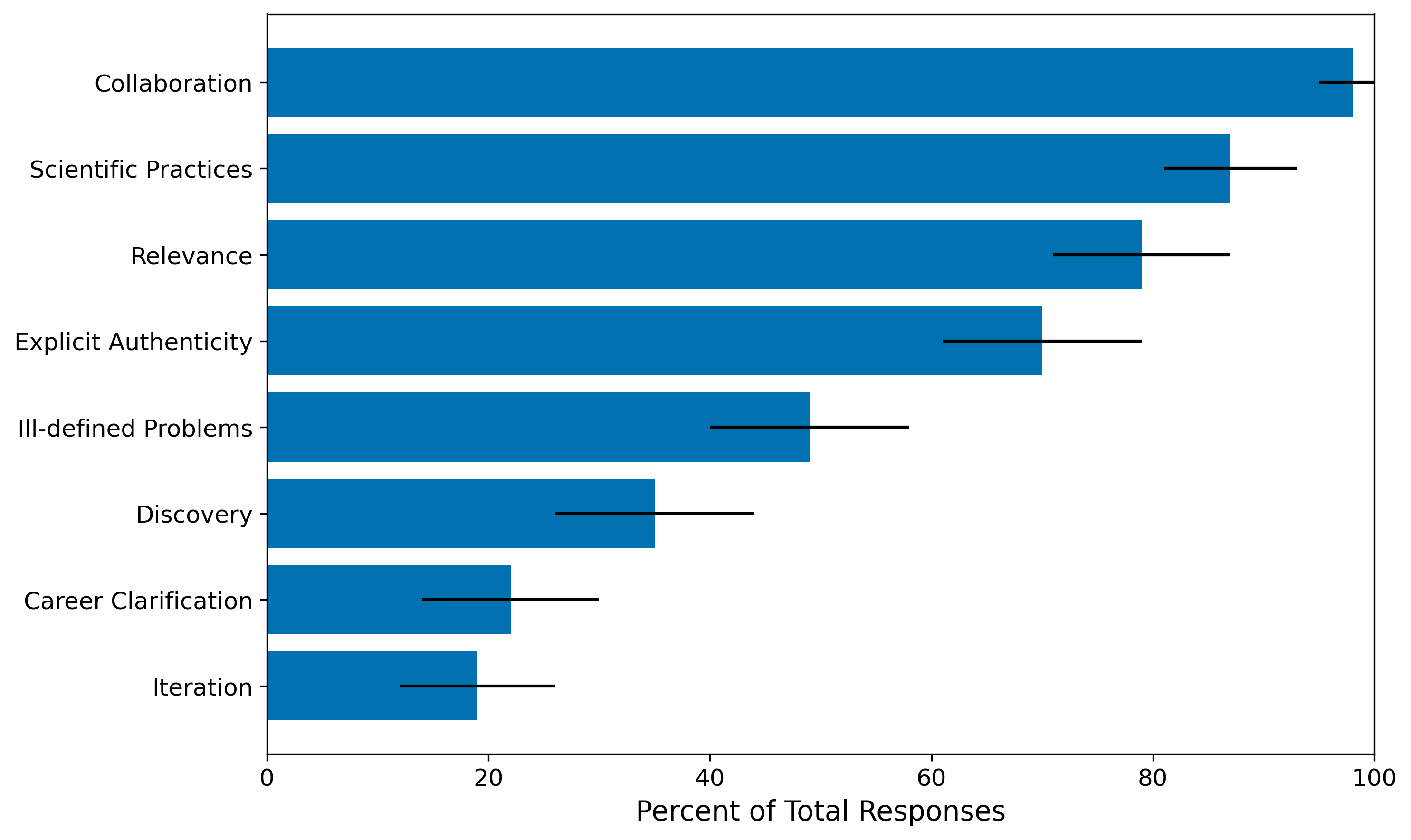}
    \caption{Frequency with which each major theme appeared across the reports, shown as the percentage of reports containing each code with 95\% confidence intervals.}
    \label{fig:cure_overview}
\end{figure*}

\begin{figure*}[t]
    \centering
    \includegraphics[width=\textwidth]{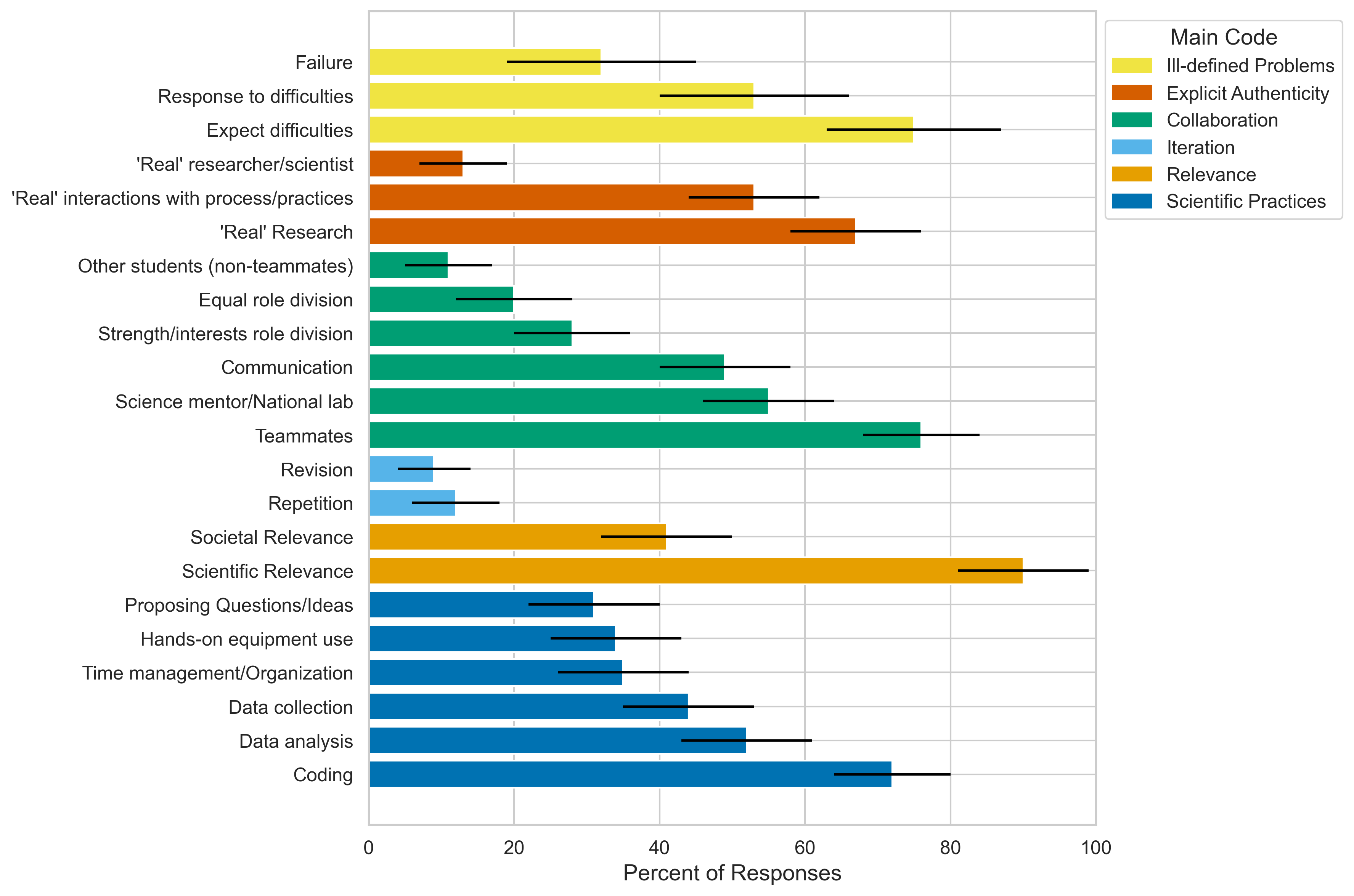}
    \caption{Occurrence of subcodes within the reports. Percentages reported for subcodes are calculated relative to the subset of reports coded under the corresponding main code, with 95\% confidence intervals shown for student-reported engagement with the subcodes of six main codes: Ill-defined Problems (yellow), Explicit Authenticity (red-orange), Collaboration (green), Iteration (sky blue), Relevance (orange), and Scientific Practices (blue). Discovery and Career Clarification do not contain subcodes and are therefore not included.}
    \label{fig:subcode_bd}
\end{figure*}

The analysis examined how students characterized their experiences with the core components of a CURE, as well as how they perceived the course as reflecting authentic research, engagement with ill-defined problems, and reflections on future academic and professional goals. In designing the course, we intentionally incorporated activities aligned with each component to ensure students would engage with a full range of authentic research practices. We present the frequency with which these themes appeared across the reports (Fig.~\ref{fig:cure_overview}), as well as the emergent subthemes associated with six of the codes (Fig.~\ref{fig:subcode_bd}). The implications of these findings are presented in the Discussion section.
\subsection{Student-identified engagement with CURE components}
To varying degrees, students discussed all five CURE components in their reports.

\subsubsection{Use of Scientific Practices}
Students brought up using scientific practices in 87\% of the reports. Scientific practices are a broad category, and all subcodes used in this work emerged from the data. It should be noted that the practices identified by students are specific to this course context, and a different set of scientific practices may emerge in other courses. Among the identified scientific practices, coding (programming) was the most commonly mentioned (72\%). Student presented the use of coding in multiple ways. Some presented it as a professional tool for scientific work:
\begin{quote}
    The most useful component of the class was the opportunity to practice python and data modeling skills. Numpy, scipy, pandas, hdf5, and any other number of numerical analysis packages that we could’ve employed on this project are extremely useful tools for any aspiring scientists and engineers to have a grasp of.
\end{quote}
Others emphasized the relevance of coding in modern research:
\begin{quote}
    We also had the opportunity to sharpen our Python skills while gathering data from repositories, which is crucial because modern research heavily relies on coding and data analysis.
\end{quote}
Even students who were not the primary `coder' in their group, still identified coding as important:
\begin{quote}
   Learning Python during the course was beneficial, even though I didn't do most of the coding. Understanding the basics of the programming helped me better engage with the data analysis process. 
\end{quote}
Students also shared that the course helped build their confidence around coding. Some students discussed being able to be creative with the code they produced. Others shared that they came into the course having no experience, but left being able to effectively use code to complete their analysis:
\begin{quote}
    Personally coming into this class I had no experience coding in Python or doing any sort of data analysis using Python, but throughout the semester I took the time to learn more about Python and by the end of the semester I was the primary contributor to all of the data analysis my group completed.
\end{quote}

Data analysis (52\%) and data collection (44\%) were the next most commonly identified scientific practices. These codes often appeared together, resulting in excerpts frequently being double coded with both subcodes, as in the following quote:
\begin{quote}
    Much of the work done in class will be with data collection, organization, and visualization. You will learn how to represent data in readable ways and how to interpret it.
\end{quote}
How students described data analysis varied. Some students explained different components of data analysis, like the quote above, and other students just said they `analyzed data.'

Time management and organization was mentioned in 35\% of the reports. Some students focused more on the time management aspect: 
\begin{quote}
    Finally, this class taught me the importance  of time management. Staying organized and keeping a consistent schedule were very essential for staying on track.
\end{quote}
And other students focused on the organization of the experimental details:
\begin{quote}
    Keep detailed records of all the observations and findings from your procedures. This makes the process easier for everyone and ensures that the research is based on solid principles. Do not forget to label all the files correctly and pay attention to every detail.
\end{quote}

Hands-on equipment use was mentioned in 34\% of the reports. Some students noted the value of working with tools not typically available in instructional labs:
\begin{quote}
    As we embarked on our experiments, we learned to handle sophisticated equipment to analyze the EQE and J-V data of solar cells. This hands-on experience was our first exposure to real-world research tools that aren't typically available in classroom settings.   
\end{quote}
Other students framed it in the context of a specific research topic:
\begin{quote}
    Something else I learned is that you want to make sure you understand every piece of equipment you are using, as there is a decent amount of machinery and tools used in this that a normal person not affiliated with this field would not really understand.
\end{quote}

Finally, 31\% of students discussed the process around their team-formulated research question. Some focused on the process of generating a question:
\begin{quote}
    Over the course of this semester, write down any questions you may have which can be answered with your data. One of these queries might become your research question at the end of the term.
\end{quote}
While some students focused on the procedural aspects of developing a research question, others connected this scientific practice to broader ideas about scientific identity and participation in authentic research.
\begin{quote}
As you progress through the class, you gain your scientist wings and are sent to devise questions to investigate using the data collected throughout the semester.
\end{quote}
Others highlighted how engaging in this scientific practice positioned their group as participants in a broader research effort:
\begin{quote}
    Each group makes their research question which they then head out to answer via using and analyzing all the data collected by each group in the class. The group gets to come up with their own methodologies, own analysis, and own conclusion, and gets to submit their research to an active researcher in solar cell technology.
\end{quote}
This response was also coded as \textit{board relevance to the scientific community} because the student framed their work as contributing to ongoing research beyond the classroom context.

\subsubsection{Relevance}
The relevance of the work done in the CURE was mentioned in 79\% of student reports. These reflections clustered into two primary categories: broad relevance to the scientific community (70\%) and relevance to society (41\%). 
Some students focused on the scientific community at the level of the science mentor and national lab:
\begin{quote}
    You will be working with actual scientific material, and creating knowledge that will be used as a baseline for the scientists who are presenting you with the prompt of your work in this course.
\end{quote}
Others situated their work within the broader field of physics:
\begin{quote}
    The research project we are helping NREL with is very important to the development of efficient energy physics.
\end{quote}
The second type of relevance students emphasized was societal impact, especially in the context of climate change and clean energy. Some focused primarily on environmental impact:
\begin{quote}
    Solar energy is abundant and renewable, and to reach a point of it being efficiently harnessed could have a very positive impact on current energy issues today. Instead of relying on nonrenewable energy sources such as coal which have large negative impacts on the environment, a clean and renewable option could be solar power.
\end{quote}
Others emphasized the impact on people and communities:
\begin{quote}
    Our class, having helped their team, was involved in something that has the potential to unlock more efficient and clean energy which could help the planet and the civilians who live here!
\end{quote}
A few students highlighted the global scale of the work:
\begin{quote}
    I think it’s important at the outset of the class to really appreciate the importance of what we’re doing; we’re contributing to a global cause.
\end{quote}
Some students described both scientific and societal relevance, and those excerpts were double-coded. For example, the following quote emphasizes both the development of new solar technologies and their potential environmental impact:
\begin{quote}
    Approach this work with seriousness, as your contributions in this class are aimed at developing a new class of solar cells that could significantly aid the green energy transition and positively impact the environment. This research is not only legitimate but could also have lasting implications for your future.
\end{quote}

\subsubsection{Discovery}
Discovery was discussed in 35\% of the reports. The majority of student-perceived discovery focused on how the specific parameters explored in the experiment would lead to the improvement of long-term stability of the PV devices.
\begin{quote}
    By identifying which factors impact the degradation of the cell the most, we can determine which parts of the cell we need to improve in development, allowing us to build a much more resilient cell.
\end{quote}
Some students took this idea a step further by discussing how discoveries related to device stability could contribute to improving the technology and, ultimately, support its commercialization, which connected discovery to relevance:
\begin{quote}
   [We], and the researchers at NREL are trying to figure out the degradation over different wavelengths and temperature. Which, in a way, will allow us to be one step closer to be able to use perovskite cells for homes in the future, especially long term.
\end{quote}

\subsubsection{Iteration}
Only 21\% of the reports mention iteration. One student gave a very general description of iteration in the course:
\begin{quote}
    We had a few hiccups in the experiment design and process this semester that required a few iterative fixes.
\end{quote}
Among the responses coded for iteration, the most common framing was repetition (57\%).
\begin{quote}
While a lot of the data collection was repetitive, it did highlight the importance of repeating data taking, especially when looking for effects over time, which is an important takeaway from this course.
\end{quote}
Another student framed repetition in the context of being mindful of any issues that may arise and need to be corrected:
\begin{quote}
   I learned to be patient. Research is not instantaneous gratification. Taking data isn’t quick and you have to be consistent week after week, even when you do streamline the data taking process, to be careful the data you are collecting is good. If you realize you did something wrong...don’t brush it under the rug otherwise you won’t be able to conduct proper analysis of the results and there won’t be any real progress. 
\end{quote}
Other students described iteration in terms of revision (43\%), typically in the context of a manufacturing defect with the first batch of cells that required iteration on stressing parameters, particularly temperature.
\begin{quote}
    For example, a lot of our cells died in the first few weeks and class and we needed to change our stressor conditions across the board due to the discovered weakness in the batch of cells we utilized.
\end{quote}
Finally, some students discussed iteration that they did not directly participate in, but was done at NLR:
\begin{quote}
    One day, all of the pixels on our PV cell died, and the research approach had to be adjusted by coating the pixels with a protective layer.
\end{quote}

\subsubsection{Collaboration}
Collaboration was by far the most discussed component in the reports, appearing in 106 of the 108 submitted assignments (98\%). Students often emphasized collaboration as a fundamental part of doing science, challenging the notion of the `lone genius' and highlighting the value of teamwork in research. For example, one student wrote:
\begin{quote}
Don’t be caught in the misconception that research is a solo effort. It takes a whole team to tackle real experimental physics research.
\end{quote}
Students also reflected on how diverse perspectives and skill sets contribute meaningfully to scientific progress:
\begin{quote}
I have learned that research not only lies in the skill of each individual, but the collaboration between individuals as a group. Working as a group made me understand my group members’ different perspectives on the research, and that has taught me a skill—that one solution is often a mix of different perspectives, and this is a core quality of collaboration that is the most important one to master.
\end{quote}
Several students noted that, prior to this course, they had few opportunities to engage in genuine collaboration. They described this CURE as a valuable experience for building practical teamwork skills:
\begin{quote}
So far in undergrad physics it is pretty likely that you haven’t had much experience working with a team on a group project, so lessons learned in this class like how to communicate with team members, how to delegate work fairly and efficiently, and how to complete a team project before a deadline will all be very valuable. These are things that will come up again and again throughout your education and career so they are important things to learn.
\end{quote}

Approximately half of the students (49\%) brought up communication as a key part of collaboration. Some student brought it up in the context of communicating in order to keep the group cohesive and on track:
\begin{quote}
    Another key skill is communication. When every group member is on the same page, everything runs smoothly. Communication is key to making sure everyone knows what is going on and what their role is.    
\end{quote}
Others focused on communication as essential for promoting fairness and accountability within the team:
\begin{quote}
    You also will be doing work  entirely  with group mates, so you need to be comfortable with communicating with them. This includes being able to resolve conflict with each other, divide up the work between each other evenly and hold them accountable for completing it, making sure everyone's opinion and analysis is valued and recorded, and treating  each  member of your team/section with respect.
\end{quote}
Conflict was a recurring theme in student reflections on communication. Some students saw communication as a tool for resolving inevitable disagreements, while others described it as a way to prevent conflict by establishing shared understanding:
\begin{quote}
    Learning how to talk to individuals with different perspectives or styles of communication is important so conflict doesn’t occur. Conflicts are destined to occur if consideration of one’s words is not taken into account.
\end{quote}
A few students also highlighted communication as a form of self-advocacy, necessary to ensure their contributions were recognized and their needs addressed:
\begin{quote}
    In the overall class a student will be successful if they advocate for themselves. If you do not contribute to the project, you will not know what is going on. If you do not speak up about you being left behind by your group, they will continue; therefore, speak up and make your opinions heard. Everyone has good ideas that should be contributed to the project.   
\end{quote}

Students also identified their collaborators in the reports. Unsurprisingly, the most frequently mentioned collaborators were their teammates (76\%). Some students acknowledged that they do not typically enjoy group work, but noted that a well-organized team dynamic helped create a more positive experience:
\begin{quote}
     I don't enjoy working in groups, but our team was well-organized, and the way we worked together felt professional, kind of like a job.   
\end{quote}
Others described forming friendships with their teammates, which they felt made it easier to divide tasks and address challenges when they arose:
\begin{quote}
    I also encourage anyone coming into this class to get to know their lab partners well. I had a fine time with mine and made some new friends. There was a lot to do together and we had to be comfortable with one another to spread the work evenly or call out someone for not doing their work.  
\end{quote}
Because students did not choose their teammates, some reflected on this as a realistic aspect of scientific research. They advised incoming participants to see it as a valuable opportunity:
\begin{quote}
    Research is very collaborative, and often you aren't working with people you already know. Working with a variety of people offers new ideas and perspectives.
\end{quote}

The science mentor and the national lab were the next most frequently mentioned collaborators (55\%). Many students emphasized that the course involved genuine collaboration with professionals in the field:
\begin{quote}
   These experiments were conducted in collaboration with the National  Renewable Energy Laboratory (NREL), a leading institution working to improve the performance, durability, and scalability of Perovskite solar cell technology. 
\end{quote}
For some students, interacting with the science mentor was one of the most meaningful aspects of the course:
\begin{quote}
    Eventually, you hear Dr. Berry's thoughts on your hard work. The best part is he acknowledges how his team plans to expand on your discoveries.    
\end{quote}
Students also recognized that this collaboration was not one-sided, that the work they completed contributed meaningfully to the national lab's ongoing research:
\begin{quote}
    We helped provide useful information to our collaborators at NREL, and you will too!  
\end{quote}
As with their team-based collaboration, students saw value in working with external researchers because it brought new perspectives and reshaped their understanding of the scientific process, recognizing that even unexpected outcomes can yield valuable information:
\begin{quote}
    An unexpected error occurred with the solar cells we were provided where overexposure to certain wavelengths destroyed them. However, the collaboration with researchers with NREL made us realize that this was still vital data that could be used by future researchers.   
\end{quote}

Other students in the course were mentioned by 11\% of participants, generally in the context of shared responsibility. Students acknowledged that their actions could affect not just their own group, but the broader research effort:
\begin{quote}
If someone takes a week of bad data, then that is one less week of data, not only you, but also other researchers (such as classmates or NREL researchers), have to work with.
\end{quote}

TAs and the instructor were mentioned by students in the reports, though generally not in the same collaborative context as teammates, the science mentor, or the national lab. TAs appeared in 25\% of the reports, typically as sources of assistance and guidance. The instructor was referenced in just 10\% of the reports, most often alongside the TAs and in relation to coursework support rather than active research collaboration.

Finally, 48\% of students discussed how roles were divided within their groups. Two recommendations emerged: giving everyone an equal opportunity to engage with all aspects of the project and dividing responsibilities based on individual preferences and strengths.

Students who favored the equal-opportunity approach (20\%) often framed it as a way for all group members to learn and contribute equally:
\begin{quote}
    I know a lot of groups split up the work so each person had the same task each week. My group rotated the tasks so everyone was exposed to all aspects of the course. I would recommend doing it this way so it doesn't get stagnant and you learn as much as possible.   
\end{quote}
Others highlighted how alternating roles helped ensure that no one felt excluded from key parts of the process:
\begin{quote}
    It can also be helpful to rotate roles each week so that everyone gets a chance to set up, follow instructions, and operate the computer. This way, you build a well-rounded understanding of the process and avoid feeling left out at any point.    
\end{quote}
Students also noted the practical importance of being familiar with multiple roles in case a teammate was absent:
\begin{quote}
    I would  also recommend being involved in multiple parts of the research project. You should  be able to take data, code, and analyze the data you get. This is particularly important if one or multiple of your group members is absent.  
\end{quote}

Students who recommended determining roles based on strengths and interests (28\%) emphasized that doing so helped the group work more efficiently:
\begin{quote}
    By dividing responsibility based on individual strengths and interests,  the team can effectively tackle different different aspects of the [work]. 
\end{quote}
Some encouraged new students to reflect on their abilities early in the course, so that they could take on tasks where they felt confident, which would, in turn, contribute to both individual and group success:
\begin{quote}
    A lesson I learned is that you need to recognize your strengths and weaknesses and how those may impact the group. You might be better at coding than at running the experiment itself, and that is okay. You should communicate this to your group members and collectively determine what roles everybody would succeed in because this will help you succeed as a group.
\end{quote}

While most students described positive collaborative experiences, a small number also discussed challenges related to uneven participation within assigned groups. The most common issue involved team members who failed to attend lab consistently or contribute to group work:

\begin{quote}
    On the other hand I know some groups that had team members consistently not show up, and that made the overall experience unenjoyable and frustrating for the other members that did show up. 
\end{quote}

Another student reflected on the importance of balancing empathy for teammates with accountability within collaborative research settings:

\begin{quote}
    In that same way, you should be comfortable with helping out your teammates and being understanding of things that come up in their lives. Despite that, don’t take advantage of people’ s kindness and don’t let your teammates take advantage of you either. We had a bit of a situation where a teammate basically never showed up to class and never contributed to assignments.
\end{quote}

Although these negative experiences were relatively uncommon, they highlight the importance of communication, accountability, and equitable participation in collaborative research environments.

\subsection{Student-identified engagement with other aspects of authentic research}
In addition to the five components of CUREs, students also discussed their perceptions of the authenticity of the research, their experiences navigating ill-defined problems, and the impact the course had their future career plans or professional goals. These themes were examined because prior work has shown that authentic research experiences can shape students' scientific identities, engagement with research, and career trajectories \cite{auchincloss_assessment_2014, brownell2015high, rowland2016we, corwin_effects_2018, werth_assessing_2022, cooper2019impact, hanauer2017, graham_persistence_2013, wilczek2022catalyzing, newell2022gains, lopatto_undergraduate_2007, harrison2011classroom}. Understanding how students interpreted these experiences provides additional insight into the ways experimental physics CUREs support student development.

\subsubsection{Explicit Authenticity}
Students shared feelings of authenticity in 70\% of the reports. Three distinct dimensions of authenticity emerged: participation in real research or science (67\%), engagement with real scientific processes or practices (53\%), and identification as real scientists or researchers (18\%). Explicit authenticity codes were assigned when students explicitly framed their experiences as `real' science, `real' research, or `real' scientific participation, rather than simply describing engagement in scientific practices. As a result, excerpts discussing activities such as data collection or collaboration were often double-coded when students also emphasized the authenticity or real-world nature of those experiences.

The most commonly identified form of explicit authenticity was students believing they had participated in real research:
\begin{quote}
This is real research, you will be collecting data that will be used in a real scientific study.
\end{quote}
Students frequently contextualized this sense of `realness' by emphasizing the impact and relevance of their contributions:
\begin{quote}
The research done in this class has a real impact! It is sent to the people working at NREL and has an actual impact on how they produce their cells.
\end{quote}
For many, explicit authenticity was also tied to experiencing the iterative and unpredictable nature of experimental work:
\begin{quote}
During my time doing this research, I learned a lot about what a real physics experiment looks like, including all the failures and unexpected outcomes that come with real research.
\end{quote}

About half of the students described `real' interactions with the scientific process. This theme followed two main threads. First, students appreciated working with authentic research equipment and following procedures that mirrored real-world experimental practices:
\begin{quote}
It is a great educational experience to work with all of the machines available to us in the lab and see what it is like conducting actual physics research, especially in a learning environment.
\end{quote}

Second, students viewed collaboration as explicitly authentic to research:
\begin{quote}
This class helped me emulate a real research setting by working with colleagues and peers to reach an overall goal.
\end{quote}

A smaller group of students described identifying as `real' scientists, researchers, or physicists. Some identified their role as both a student and a researcher:
\begin{quote}
    I was a student, as well as a researcher, in this class when it was first redesigned.
\end{quote}
Other students viewed their role as a researcher as new. A majority of students in the CURE have not participated in traditional research experiences prior to their enrollment in the course, so this is understandable:
\begin{quote}
    As a new researcher, I found much of the process to be new, but it was a challenge that I welcomed.
\end{quote}
Several students described the course as positively impacting their emerging identities as scientists:
\begin{quote}
    As a new physicist, being able to contribute to the scientific community in such a direct and tangible way was very inspiring.
\end{quote}
A few students also reflected on the broader implications of participation, positioning themselves within a larger community of researchers:
\begin{quote}
    As a student in Physics 2150 you are not just an undergraduate in the CU Physics department, you are a member of an international research effort; your responsibility is no longer just to yourself, but rather to the scientific community as a whole.
\end{quote}

\subsubsection{Ill-defined Problems}\label{illdefined}
In the instructor interviews conducted in the fall of 2023, some instructors expressed concern that students would not have the `academic maturity' to positively engage with the inherent messiness of experimental research \cite{merritt2024}. However, we found that nearly half of students (49\%) discussed their experiences with ill-defined problems and how they responded to them.

In 30\% of the reports, students shared that difficulties are an expected part of doing research and that this is a normal, and even valuable, aspect of the process:
\begin{quote}
    Finally, know that not everything will go to plan. We learned this when half way through the semester nearly all of our solar cells ended up degrading much more than expected, causing them to die entirely. This meant we couldn’t take any more data on them, throwing a big wrench in the middle of the research. However, this was a valuable lesson in real experimental physics research. In real research, not everything will always go to plan—and that’s ok.
\end{quote}
Some students emphasized that these challenges are not just expected but can help formulate new research questions and directions:
\begin{quote}
    There is not necessarily a set schedule or lesson plan, as much as we can try to implement one, sometimes things go wrong and plans must change. Although, as someone who plans out every hour of every day, this can be frustrating, I appreciated getting a real glimpse into what research entails, realistically. Research is not a linear process and many times you will end the semester with more questions than you had when you came in. But this is okay and this is a fundamental part of learning new things!
\end{quote}
Others were more direct in their messaging regarding this theme:
\begin{quote}
    Take your mistakes and learn from them, knowing that you will make more and you will learn from those as well!
\end{quote}

In 37\% of the reports, students gave advice on how to respond to unexpected outcomes. The most common recommendation was perseverance. Students shared both specific examples and broader encouragement to not let setbacks derail the research process:
\begin{quote}
One important  lesson that I learned during the data collection process was how to deal with unexpected results. Early on in the process our cell, which was intended to last the whole semester, completely died which caused some concern. However, it turned out that this was a common problem that many groups were facing and it was most likely due to something unexpected in the manufacturing  process. We were quickly given a new cell and continued on with our data collection. It is important to not let minor setbacks like this get in the way of the broader research process.
\end{quote}
Students acknowledged that these moments can be discouraging, but encouraged future participants to persist:
\begin{quote}
    Firstly, you are bound to fail during your research experience (whether it is your fault or not is not important; what is important is that it is bound to happen) and this is going to make you want to quit, to which I say try your best. All you can do after a failure is to pick yourself up, dust yourself off and continue onwards.
\end{quote}
Others discussed reframing their understanding of failure in science altogether:
\begin{quote}
The most valuable lesson I learned from this project is that there is no such thing as failing in research. My cell died in the experiment and I was sad but my group and I moved on to a new cell and continued the experiment. It’s hard to realize in science that there is no failing, but it just didn't work for that set of boundaries placed for that object. It's important to continue and not get discouraged by the results.
\end{quote}

Another commonly recommended response was to ask questions of teammates, TAs, or the instructor:
\begin{quote}
    Don't be afraid to ask questions. Nobody expects you to know everything, and you'll learn a lot just by speaking up.
\end{quote}

Failure was explicitly mentioned in 32\% of the reports. However, all instances were also double coded under `Response to Difficulties,' reinforcing our interpretation that students did not view failure as an endpoint. Instead, students tended to frame failure in terms of how they responded to setbacks, including adapting plans, seeking support, and continuing the research process:
\begin{quote}
    Continuing onwards, failure is bound to happen in life (especially in research), so learning how to fail and how to bounce back from failure is something that I am very grateful I learned in this course
\end{quote}
This coding overlap may suggest that students did not typically conceptualize failure as separate from the practice of doing research. Rather, failure was often framed as something to be navigated through persistence, troubleshooting, and collaboration.

\subsubsection{Reflections on Future Academic and Professional Goals}
Reflections on future academic and professional goals appeared in 22\% of the reports. Students who discussed this theme reflected on how participation in the course informed their future academic and professional pathways. These responses capture students' immediate reflections at the conclusion of the course rather than long-term career outcomes.

Some student focused on how the course prepared them for future academic goals:
\begin{quote}
    This research contributed to my future goals of grad school and working in a lab by giving me research experience (which will be a large part of my future) and teaching me about the research process.
\end{quote}
Others emphasized the course helped to prepare them for employment opportunities:
\begin{quote}
    The most valuable takeaway from the class was just getting the opportunity to engage in actual physics research, so now I have that experience that I can take with me when trying to get an internship or job.
\end{quote}
For some students, the course may not have pushed them in a particular direction, but provided an experience that will help them make more informed choices in the future:
\begin{quote}
    I don't believe that I have more desire to find an occupation in an experimental job, but I am far more interested in the field as a whole, so it's definitely still a consideration.
\end{quote}

\section{Discussion}
In this section, we provide an interpretation of the qualitative findings and consider their implications for the design of experimental physics CUREs. The instructional suggestions discussed throughout this section represent implications arising from the present findings rather than interventions whose effectiveness was evaluated in this study. Examining their impact will require future investigation. We first discuss how students perceived the core CURE components integrated into the course (RQ1), followed by how students described their authentic research experiences, including authenticity, engagement with ill-defined problems (RQ2), and their reflections on future academic and professional goals (RQ3).

\subsection{RQ1: Student perceptions of the designed CURE components}
\subsubsection{Use of Scientific Practices}
On the whole, the course successfully engaged students with scientific practices. The student reflections showed strong engagement with several course learning goals, particularly coding/data analysis and research question development. It was also encouraging to see a third of the students identify the importance of time management and organization. Prior work in STEM and physics education has emphasized the importance of professional skills such as documentation, organization, and time management for both research and workforce preparation (e.g., \cite{leak2018,karimi2021strategically,alicea-munoz2021}). Given the amount of time students spent using the J-V and EQE apparatus throughout the course, it was interesting that equipment use was not discussed more often. Students worked directly with the apparatus for half of the semester and had one full lecture dedicated to the EQE apparatus.  Future iterations of the course may benefit from additional opportunities for students to reflect on the role of instrumentation in the research process.

\subsubsection{Relevance}
With nearly 80\% of students discussing the relevance of the project in their reports, this component appears to be well-integrated into the course experience. This marks a notable shift from the pre-transformation version of the course, in which 65\% of students reported feeling that the work they were doing lacked relevance beyond the classroom \cite{micahthesis}.

This suggests the instructor was successful in framing the `why' of the research from the outset of the course. In-person lab meetings with the science mentor further reinforced this framing, helping to solidify students' understanding of the project's significance. In the case of this CURE, the topic of solar energy provided a clear bridge between classroom research and real-world impact. An equally important aspect of relevance, however, was that students' data contributed to an ongoing research collaboration with the NLR. Knowing that their results would be used by an external research partner reinforced that the results of their work extended beyond the classroom. However, this may not generalize across all possible research topics. For courses focused on less obviously applied research topics, it may be important to build in more explicit reminders or activities that contextualize the relevance of the research topic, while also clearly communicating how students' work contributes to ongoing efforts outside the classroom.

\subsubsection{Discovery}
A third of students mentioned discovery in their reports. This does not indicate that students did not experience discovery in the course, but rather that discovery was less salient than some of the other CURE components in this particular reflection artifact. The framing of the assignment, which did not explicitly prompt students to reflect on novelty or knowledge generation. Corwin et al., 2018 found that discovery was not significantly or positively related to students’ emotional ownership of their work \cite{corwin_effects_2018}, suggesting that discovery, while important, is not always the most personally meaningful component of a CURE. In the context of these reports, written as letters to future participants and intended to offer encouragement and advice, discovery was not a primary focus. Instead, students more often emphasized the relevance of the project. Future iterations of the course may benefit from more explicitly highlighting students' role in generating new scientific knowledge and providing opportunities for students to reflect on how their work contributes to discovery. To better understand how students engage with the discovery component, future work should examine additional course artifacts to determine whether discovery is less prominent in the overall course experience or simply not emphasized in this particular assignment.

\subsubsection{Iteration}
Iteration was discussed less frequently than other CURE components in the reports, despite the course intentionally incorporating two key opportunities for iteration. The first involved the template code provided to students in weekly Colab notebooks. Each week’s analysis built upon skills developed in previous weeks, and by the final third of the semester, students adapted and extended this code as the foundation for their team-designed analysis. The second area of iteration occurred during the development of students' team-formulated research questions. Students submitted a written proposal, received detailed feedback, and revised their work accordingly. These built-in structures provided students opportunities to develop and revise research components. However, students may not have recognized these activities as forms of iteration. Alternatively, while students experienced iteration throughout the course, they may not have viewed it as one of the most memorable or important aspects of the research experience to communicate to future students. As with the discovery component, iteration was discussed less frequently within this particular assignment than other aspects of students' experiences, such as navigating ill-defined problems. To better understand how students perceive and engage with iteration, we plan to examine additional course artifacts in future analyses. We also recommend integrating more opportunities for peer feedback, which may make iteration more visible and meaningful. This could be done formally through peer review of proposals or other documents, or more informally by structuring time in lab sessions to include short presentations and feedback exchanges between teams.

\subsubsection{Collaboration}
With nearly every student mentioning collaboration in their reports, the collaboration component was effectively integrated into the course, which was associated with one of the main learning goals of the course.  Students were assigned to teams rather than selecting their own. While instructors may be concerned that assigned teams could lead to negative student experiences, this was not broadly observed in this course. Within the context of the reports, even students describing frustrating collaborative experiences often framed them as opportunities to reflect on what makes an effective scientific collaborator.

While it was expected that students would identify their teammates as collaborators, it was encouraging to see that more than half of the students also recognized the science mentor and national lab as active collaborators. It was less encouraging to note that other classmates were infrequently described in this way. Students tended to view TAs and the instructor primarily as course support or authority figures rather than as fellow researchers. As the pool of TAs who have facilitated the CURE continues to grow, we plan to conduct a set of interviews to explore whether TAs also perceive their roles as distinct from that of collaborators, and to identify practices that might help shift the TA role from facilitator to co-researcher. Given the size of the course (75–-150 students per semester) and the number of lab sections per semester (6–-14), integrating the instructor into a more collaborative role is particularly challenging. This may be less of a limitation in smaller courses. However, it may also reflect an unavoidable dynamic. The instructor and TAs' role in assigning grades may inherently prevent students from viewing them as collaborators in the same way as the other groups mentioned.

\subsection{RQ2: Student interpretations of authentic research experiences}
\subsubsection{Explicit Authenticity}
A majority of students acknowledged explicit authenticity of the research conducted in the course. Their views on the relevance of their work appeared to shape their perception of its authenticity. Even students who approached the experience through a course-based lens, rather than a research-focused one, still described participating in authentic scientific processes and practices. However, a much smaller portion of students identified themselves as authentic researchers or scientists. This may be related to the structure of the course. While students generated real data, their results are sent to external collaborators at the national lab, who will synthesize the findings. As a result, students may have felt they contributed to real research, but not that they filled the role of a researcher or scientist. In future semesters, it may be helpful to more explicitly emphasize that `real' researchers play a variety of roles in the scientific process and that all of those roles are valuable. Reinforcing this message throughout the course could support students in seeing themselves as part of the broader research community. Additionally, creating opportunities for students to share the outcomes of their work, such as through an end-of-semester departmental poster symposium, could further support the development of their identity as `real' scientists.

\subsubsection{Ill-defined Problems}
With half of the students discussing ill-defined problems in their reports, we propose that students may be more capable of engaging productively with ambiguity than some instructors expect. Students described experiencing device failure, changing plans, and unexpected complications, while also recognizing these challenges as a normal part of the research process. Although students acknowledged that these moments could be difficult and frustrating, they frequently emphasized the importance of perseverance, adapting plans, and seeking help as part of navigating ill-defined problems in science. Students also highlighted the importance of relying on teammates and asking questions of the TAs and instructor when responding to these situations.

These findings suggest that explicitly normalizing setbacks and uncertainty early in the course may help students engage more productively with ill-defined problems. Prior work has shown that students' responses to setbacks and uncertainty in undergraduate research environments are shaped by both instructional context and students' perceptions of challenges and failures (e.g., \cite{elbenzayas2016, henry_fail_2019, corwin2022students, corwin2022confronting}). Supportive instructional environments can promote more challenge-engaging responses to ill-defined problems, helping students interpret setbacks as a normal component of authentic scientific work rather than as evidence of inability or failure. During this course, the instructor intentionally framed setbacks as a normal part of research from the very first class meeting. Additionally, when the course experienced significant device failure after the first week of stressing, the instructor immediately addressed the issue and provided students with updates as the situation was being remedied. The TAs also played an important role in shaping the tone of the lab environment through their responses to unexpected issues during lab sessions. In future semesters, it will be important to continue establishing strong instructional and social supports to help students navigate the ambiguity inherent in authentic research experiences.

\subsection{RQ3: Reflections on Future Academic and Professional Goals}
Although reflections on future academic and professional goals were not a major theme in the reports, students who discussed them described the course as helping them consider both academic and industry pathways and reflect on future decisions. These findings capture students' immediate reflections at the conclusion of the course rather than long-term career outcomes or retention. Given that most students were in their second year of study, these relatively infrequent reflections may reflect students' early stage of career exploration rather than a lack of impact. Currently, the course includes only a brief discussion of traditional undergraduate research opportunities during meetings with the science mentor and no dedicated career exploration activities. These findings suggest that intentionally incorporating discussions of career pathways or professional development into the course may help more students connect their research experiences to future academic and professional goals.

\section{Summary and Future Work}\label{conclusions}
Analysis of the \textit{Report to Future Researchers} reflections suggests that students experienced the CURE in ways that closely aligned with its intentional design, while also revealing differences in which aspects of the research experience were most salient to students. With respect to RQ1, students described experiences associated with all five core CURE components integrated into the course, with collaboration, relevance, and scientific practices appearing most prominently in their reflections. Discovery and iteration were discussed less frequently. These differences do not indicate whether the students experienced or learned individual components to different degrees. Rather, they identify which aspects of the research experience students most often chose to emphasize when communicating their experience to future participants. Together, these findings show alignment between the intentional design of the course and how students described their experiences. Students described experiences associated with all five CURE components without being explicitly prompted about the individual components, while differences in their relative salience highlight areas that may benefit from additional instructional emphasis or opportunities for reflection.

 With respect to RQ2, students' descriptions of authentic research extended beyond participation in particular scientific practices. Students associated authenticity with making meaningful scientific contributions, engaging with multiple scientific practices, and navigating the uncertainty and setbacks inherent in authentic research. At the same time, although many students characterized the work as authentic, fewer explicitly identified themselves as researchers or scientists. This distinction suggests that participating in and recognizing authentic scientific work does not necessarily translate immediately into seeing oneself as a member of the scientific community, highlighting a potential role for instructional structures that make students' contributions and positions within the broader research endeavor more visible. 
 
 RQ3 provided an additional finding beyond those captured by the CURE and authenticity frameworks. Some students described the course as informing their immediate thinking about future academic and professional pathways. These reflections do not establish longer-term effects on retention or career decisions, but they demonstrate that, for some students, participation in the CURE was connected to their immediate thinking about what they might pursue next. This finding suggests an opportunity to investigate more directly how early research experiences interact with students' developing academic and professional goals. 
 
 Overall, these finding illustrate how students experienced multiple dimensions of authentic research within an experimental physics CURE. The five CURE components examined through RQ1 capture student recognition and characterization of intentionally designed features of the course, while RQ2 examines how students interpreted authenticity and engagement with ill-defined problems during their research experience. RQ3 captures how some students connected their experience to their immediate thinking about future pathways. Table~\ref{tab:summary} summarizes the these findings across the three research questions and identifies the corresponding considerations for designing an experimental physics CURE. 

\begin{table*}[t]
\caption{Summary of the principal findings across the three research questions and corresponding design considerations for experimental physics CUREs.}
\label{tab:summary}
\centering
\small
\begin{tabularx}{\textwidth}{p{0.18\textwidth}X X}
\toprule
\textbf{Component} & \textbf{Primary Finding} & \textbf{Implications for Course Design} \\
\midrule
\multicolumn{3}{l}{\textbf{RQ1: CURE Components} (Auchincloss et al. \cite{auchincloss_assessment_2014})} \\
\midrule
Scientific Practices &
Students frequently recognized coding, data analysis, and research question development as central components of scientific practice. &
Students may benefit from explicit opportunities to reflect on the role of experimental instrumentation as part of scientific practice. \\

Relevance &
Students consistently recognized the scientific and societal relevance of the project. &
Early framing and interactions with external science mentors may help students connect course activities to broader scientific and societal impacts. \\

Discovery &
Students discussed discovery less frequently than other CURE components in the reports. &
Students may benefit from explicit opportunities to reflect on their role in generating new scientific knowledge. \\

Iteration &
Students discussed iteration less frequently than other CURE components despite multiple intentionally designed opportunities for iteration. &
Structured opportunities for revision and peer feedback may help make iteration more visible to students. \\

Collaboration &
Students viewed teammates and the external science mentor as collaborators, but viewed TAs and instructors primarily as facilitators or authority figures. &
Students may benefit from instructional structures that emphasize the collaborative roles of mentors, TAs, and other members of the research community. \\

\addlinespace
\multicolumn{3}{l}{\textbf{RQ2: Authentic Research Experiences} (Goodwin et al. \cite{goodwin2021science}; Oliver et al. \cite{oliver2023})} \\
\midrule
Explicit Authenticity &
Students described authentic research as contributing to meaningful scientific work and engaging in authentic scientific practices, but fewer identified themselves as researchers or scientists. &
Helping students recognize that meaningful scientific contributions occur through many roles within a research team may strengthen students' identification as researchers. \\

Ill-defined Problems &
Students viewed setbacks as a normal part of research and emphasized perseverance, adaptation, and seeking support when navigating uncertainty. &
Explicitly normalizing uncertainty early in the course may help students engage productively with ambiguity. \\

\addlinespace
\multicolumn{3}{l}{\textbf{RQ3: Academic and Professional Goals}} \\
\midrule

Reflections on Future Academic and Professional Goals &
Some students reflected on how the course informed their immediate academic and professional thinking. &
Providing opportunities for career exploration and professional reflection may help students connect research experiences to future academic and professional pathways. \\

\bottomrule
\end{tabularx}
\end{table*}

While the dimensions of students' experiences examined across all three research questions were analyzed separately, student reflections frequently demonstrated overlap among them. These findings suggest that students may experience different dimensions of a CURE not as isolated constructs, but as interconnected aspects of participation in authentic scientific work. This interpretation is consistent with sociocultural perspectives of learning, which emphasize that learning and identity development are shaped through engagement with disciplinary practices and participation in scientific communities.

These findings also provide insight into the development and initial implementation of the course itself. This work demonstrates one approach to integrating student participation in an ongoing external research collaboration and the defining features of a CURE into an in-person experimental physics lab course. The initial implementation shows that students recognized many of the intentionally designed course themes in their own research experience and can engaged with authentic elements of scientific work, including collaboration, the broader relevance of their work, and the uncertainty associated with authentic research. At the same time, differences in what the students emphasized reveals areas where intentional course design alone may not make particular aspects of the research process equally visible or meaningful to students. Documenting both the course model and students' experiences within it provides connected information for instructors considering how authentic research experiences might be implemented and supported in experimental physics lab courses.

In addition to informing broader considerations for experimental physics CURE design, the findings from this first implementation suggest several opportunities to further refine this course. As we prepare for future iterations, we are considering the following areas for improvement:
\begin{itemize}
    \item Increasing opportunities for students to reflect on the role of instrumentation in the research process
    \item Highlighting the role students play not only in collecting data, but in generating new knowledge to address a novel scientific question
    \item Revising assignments and in-lab activities to encourage cross-team engagement and create opportunities for peer review
    \item Strengthening the role of the TA as a fellow researcher and scientific collaborator
    \item Exploring ways to integrate more career exploration and professional development into the course
\end{itemize}

Future work will examine students' experiences with individual CURE components and other elements of authentic research drawing on additional course artifacts, such as weekly reflections and survey responses. Unlike the reports, which offer a broad overview of the student experience, these additional artifacts are more narrowly focused on specific components and aspects of authentic research. Analyzing these sources will provide additional perspective on students' experiences and may help distinguish aspects of the course that were less salient in the reports from those that warrant further attention in future course design.

On a large scale, this work contributes to a growing body of evidence-based practices for designing and implementing CUREs in physics by examining both the development and initial implementation of an in-person experimental physics CURE and students' experiences within it. As physics departments seek to expand research opportunities for undergraduates, particularly in the early stages of their academic careers, it will be important to provide resources to support the creation of scalable, sustainable CUREs that promote meaningful engagement with the scientific process.

\section{Acknowledgments}
This work is supported by the NSF Mathematical and Physical Sciences Ascending Postdoctoral Research Fellowship program (PHY 2316504) and PHY 2317149. We would like to thank the science mentor for the course, Joe Berry. We would also like to thank Kristopher Bunker, Bethany Wilcox, Micah Kretchmer, Rob Tirawat, and Rosie Bramante for their contributions to the development and implementation of the CURE. Colleagues, Shams El-Adawy, Kristin Oliver, and Qiaoyi (Joey) Liu, are also thanked for their insightful conversations regarding this research. Finally, we would like to thank the student participants for their efforts and contributions in the course and to photovoltaics research.

\bibliography{references}

@book{gentile_undergraduate_2017,
	address = {Washington, D.C.},
	title = {Undergraduate {Research} {Experiences} for {STEM} {Students}: {Successes}, {Challenges}, and {Opportunities}},
	isbn = {978-0-309-45280-9},
	shorttitle = {Undergraduate {Research} {Experiences} for {STEM} {Students}},
	url = {https://www.nap.edu/catalog/24622},
	language = {english},
	urldate = {2023-01-11},
	publisher = {National Academies Press},
	author = {{Committee on Strengthening Research Experiences for Undergraduate STEM Students} and {Board on Science Education} and {Division of Behavioral and Social Sciences and Education} and {Board on Life Sciences} and {Division on Earth and Life Studies} and {National Academies of Sciences, Engineering, and Medicine}},
	editor = {Gentile, James and Brenner, Kerry and Stephens, Amy},
	month = may,
	year = {2017},
	doi = {10.17226/24622},
}

@article{auchincloss_assessment_2014,
	title = {Assessment of {Course}-{Based} {Undergraduate} {Research} {Experiences}: {A} {Meeting} {Report}},
	volume = {13},
	issn = {1931-7913},
	shorttitle = {Assessment of {Course}-{Based} {Undergraduate} {Research} {Experiences}},
	url = {https://www.lifescied.org/doi/10.1187/cbe.14-01-0004},
	doi = {10.1187/cbe.14-01-0004},
	language = {english},
	number = {1},
	urldate = {2023-01-11},
	journal = {CBE—Life Sciences Education},
	author = {Auchincloss, Lisa Corwin and Laursen, Sandra L. and Branchaw, Janet L. and Eagan, Kevin and Graham, Mark and Hanauer, David I. and Lawrie, Gwendolyn and McLinn, Colleen M. and Pelaez, Nancy and Rowland, Susan and Towns, Marcy and Trautmann, Nancy M. and Varma-Nelson, Pratibha and Weston, Timothy J. and Dolan, Erin L.},
	month = mar,
	year = {2014},
	pages = {29--40},
}

@article{bangera_course-based_2014,
	title = {Course-{Based} {Undergraduate} {Research} {Experiences} {Can} {Make} {Scientific} {Research} {More} {Inclusive}},
	volume = {13},
	issn = {1931-7913},
	url = {https://www.lifescied.org/doi/10.1187/cbe.14-06-0099},
	doi = {10.1187/cbe.14-06-0099},
	language = {english},
	number = {4},
	urldate = {2023-01-11},
	journal = {CBE—Life Sciences Education},
	author = {Bangera, Gita and Brownell, Sara E.},
	editor = {Hatfull, Graham},
	month = dec,
	year = {2014},
	pages = {602--606},
}

@article{werth_impacts_2022,
	title = {Impacts on student learning, confidence, and affect in a remote, large-enrollment, course-based undergraduate research experience in physics},
	volume = {18},
	issn = {2469-9896},
	url = {https://link.aps.org/doi/10.1103/PhysRevPhysEducRes.18.010129},
	doi = {10.1103/PhysRevPhysEducRes.18.010129},
	language = {english},
	number = {1},
	urldate = {2023-01-11},
	journal = {Physical Review Physics Education Research},
	author = {Werth, Alexandra and West, Colin G. and Lewandowski, H. J.},
	month = apr,
	year = {2022},
	pages = {010129},
}

@article{werth_assessing_2022,
	title = {Assessing student engagement with teamwork in an online, large-enrollment course-based undergraduate research experience in physics},
	volume = {18},
	issn = {2469-9896},
	url = {https://link.aps.org/doi/10.1103/PhysRevPhysEducRes.18.020128},
	doi = {10.1103/PhysRevPhysEducRes.18.020128},
	language = {english},
	number = {2},
	urldate = {2023-01-11},
	journal = {Physical Review Physics Education Research},
	author = {Werth, Alexandra and Oliver, Kristin and West, Colin G. and Lewandowski, H. J.},
	month = oct,
	year = {2022},
	pages = {020128},
}

@article{linn_undergraduate_2015,
	title = {Undergraduate research experiences: {Impacts} and opportunities},
	volume = {347},
	issn = {0036-8075, 1095-9203},
	shorttitle = {Undergraduate research experiences},
	url = {https://www.science.org/doi/10.1126/science.1261757},
	doi = {10.1126/science.1261757},
	language = {english},
	number = {6222},
	urldate = {2023-01-11},
	journal = {Science},
	author = {Linn, Marcia C. and Palmer, Erin and Baranger, Anne and Gerard, Elizabeth and Stone, Elisa},
	month = feb,
	year = {2015},
	pages = {1261757},
}

@article{lopatto_survey_2004,
	title = {Survey of {Undergraduate} {Research} {Experiences} ({SURE}): {First} {Findings}},
	volume = {3},
	issn = {1536-7509},
	shorttitle = {Survey of {Undergraduate} {Research} {Experiences} ({SURE})},
	url = {https://www.lifescied.org/doi/10.1187/cbe.04-07-0045},
	doi = {10.1187/cbe.04-07-0045},
	language = {english},
	number = {4},
	urldate = {2023-01-11},
	journal = {Cell Biology Education},
	author = {Lopatto, David},
	month = dec,
	year = {2004},
	pages = {270--277},
}

@article{thiry_what_2011,
	title = {What {Experiences} {Help} {Students} {Become} {Scientists}? {A} {Comparative} {Study} of {Research} and other {Sources} of {Personal} and {Professional} {Gains} for {STEM} {Undergraduates}},
	volume = {82},
	issn = {0022-1546, 1538-4640},
	shorttitle = {What {Experiences} {Help} {Students} {Become} {Scientists}?},
	url = {https://www.tandfonline.com/doi/full/10.1080/00221546.2011.11777209},
	doi = {10.1080/00221546.2011.11777209},
	language = {english},
	number = {4},
	urldate = {2023-01-11},
	journal = {The Journal of Higher Education},
	author = {Thiry, Heather and Laursen, Sandra L. and Hunter, Anne-Barrie},
	month = jul,
	year = {2011},
	pages = {357--388},
}

@article{hunter_becoming_2007,
	title = {Becoming a scientist: {The} role of undergraduate research in students' cognitive, personal, and professional development},
	volume = {91},
	issn = {00368326, 1098237X},
	shorttitle = {Becoming a scientist},
	url = {https://onlinelibrary.wiley.com/doi/10.1002/sce.20173},
	doi = {10.1002/sce.20173},
	language = {english},
	number = {1},
	urldate = {2023-01-11},
	journal = {Science Education},
	author = {Hunter, Anne-Barrie and Laursen, Sandra L. and Seymour, Elaine},
	month = jan,
	year = {2007},
	pages = {36--74},
}

@article{lopatto_undergraduate_2007,
	title = {Undergraduate {Research} {Experiences} {Support} {Science} {Career} {Decisions} and {Active} {Learning}},
	volume = {6},
	issn = {1931-7913},
	url = {https://www.lifescied.org/doi/10.1187/cbe.07-06-0039},
	doi = {10.1187/cbe.07-06-0039},
	language = {english},
	number = {4},
	urldate = {2023-01-11},
	journal = {CBE—Life Sciences Education},
	author = {Lopatto, David},
	editor = {Williams, Paul},
	month = dec,
	year = {2007},
	pages = {297--306},
}

@article{trott_exploring_2020,
	title = {Exploring the long-term academic and career impacts of undergraduate research in geoscience: {A} case study},
	volume = {68},
	issn = {1089-9995, 2158-1428},
	shorttitle = {Exploring the long-term academic and career impacts of undergraduate research in geoscience},
	url = {https://www.tandfonline.com/doi/full/10.1080/10899995.2019.1591146},
	doi = {10.1080/10899995.2019.1591146},
	language = {english},
	number = {1},
	urldate = {2023-01-11},
	journal = {Journal of Geoscience Education},
	author = {Trott, Carlie D. and Sample McMeeking, Laura B. and Bowker, Cheryl L. and Boyd, Kathryn J.},
	month = jan,
	year = {2020},
	pages = {65--79},
}

@article{seymour_establishing_2004,
	title = {Establishing the benefits of research experiences for undergraduates in the sciences: {First} findings from a three-year study},
	volume = {88},
	issn = {0036-8326, 1098-237X},
	shorttitle = {Establishing the benefits of research experiences for undergraduates in the sciences},
	url = {https://onlinelibrary.wiley.com/doi/10.1002/sce.10131},
	doi = {10.1002/sce.10131},
	language = {english},
	number = {4},
	urldate = {2023-01-13},
	journal = {Science Education},
	author = {Seymour, Elaine and Hunter, Anne-Barrie and Laursen, Sandra L. and DeAntoni, Tracee},
	month = jul,
	year = {2004},
	pages = {493--534},
}

@article{buchanan2022,
author = {Buchanan, Alaina J. and Fisher, Ginger R.},
title = {Current Status and Implementation of Science Practices in Course-Based Undergraduate Research Experiences (CUREs): A Systematic Literature Review},
journal = {CBE—Life Sciences Education},
volume = {21},
number = {4},
pages = {ar83},
year = {2022},
doi = {10.1187/cbe.22-04-0069},
    note ={PMID: 36318310},

URL = { 
    
        https://doi.org/10.1187/cbe.22-04-0069
    
    

},
eprint = { 
    
        https://doi.org/10.1187/cbe.22-04-0069
    
    

}
}

@article{henry_fail_2019,
	title = {\textit{{FAIL}} {Is} {Not} a {Four}-{Letter} {Word}: {A} {Theoretical} {Framework} for {Exploring} {Undergraduate} {Students}’ {Approaches} to {Academic} {Challenge} and {Responses} to {Failure} in {STEM} {Learning} {Environments}},
	volume = {18},
	issn = {1931-7913},
	shorttitle = {\textit{{FAIL}} {Is} {Not} a {Four}-{Letter} {Word}},
	url = {https://www.lifescied.org/doi/10.1187/cbe.18-06-0108},
	doi = {10.1187/cbe.18-06-0108},
	language = {english},
	number = {1},
	urldate = {2023-01-17},
	journal = {CBE—Life Sciences Education},
	author = {Henry, Meredith A. and Shorter, Shayla and Charkoudian, Louise and Heemstra, Jennifer M. and Corwin, Lisa A.},
	editor = {Gardner, Stephanie},
	month = mar,
	year = {2019},
	pages = {ar11},
}

@article{eagan2013,
author = {M. Kevin Eagan, Jr. and Sylvia Hurtado and Mitchell J. Chang and Gina A. Garcia and Felisha A. Herrera and Juan C. Garibay},
title ={Making a Difference in Science Education: The Impact of Undergraduate Research Programs},
journal = {American Educational Research Journal},
volume = {50},
number = {4},
pages = {683-713},
year = {2013},
doi = {10.3102/0002831213482038},
    note ={PMID: 25190821},

URL = { 
        https://doi.org/10.3102/0002831213482038
    
},
eprint = { 
        https://doi.org/10.3102/0002831213482038
    
}
}

@article{pierszalowski_systematic_2021,
	title = {A {Systematic} {Review} of {Barriers} to {Accessing} {Undergraduate} {Research} for {STEM} {Students}: {Problematizing} {Under}-{Researched} {Factors} for {Students} of {Color}},
	volume = {10},
	issn = {2076-0760},
	shorttitle = {A {Systematic} {Review} of {Barriers} to {Accessing} {Undergraduate} {Research} for {STEM} {Students}},
	url = {https://www.mdpi.com/2076-0760/10/9/328},
	doi = {10.3390/socsci10090328},
	language = {english},
	number = {9},
	urldate = {2023-01-18},
	journal = {Social Sciences},
	author = {Pierszalowski, Sophie and Bouwma-Gearhart, Jana and Marlow, Lindsay},
	month = sep,
	year = {2021},
	pages = {328},
}

@article{jones2010,
author = {Melanie T. Jones and Amy E. L. Barlow and Merna Villarejo},
title = {Importance of Undergraduate Research for Minority Persistence and Achievement in Biology},
journal = {The Journal of Higher Education},
volume = {81},
number = {1},
pages = {82-115},
year  = {2010},
publisher = {Routledge},
doi = {10.1080/00221546.2010.11778971},

URL = { 
    
        https://doi.org/10.1080/00221546.2010.11778971
    
    

},
eprint = { 
    
        https://doi.org/10.1080/00221546.2010.11778971
    
    

}

}

@article{carlone2007,
author = {Carlone, Heidi B. and Johnson, Angela},
title = {Understanding the science experiences of successful women of color: Science identity as an analytic lens},
journal = {Journal of Research in Science Teaching},
volume = {44},
number = {8},
pages = {1187-1218},
doi = {https://doi.org/10.1002/tea.20237},
url = {https://onlinelibrary.wiley.com/doi/abs/10.1002/tea.20237},
eprint = {https://onlinelibrary.wiley.com/doi/pdf/10.1002/tea.20237},
year = {2007}
}

@article{pierszalowski_overcoming_2018,
	title = {Overcoming {Barriers} in {Access} to {High} {Quality} {Education} {After} {Matriculation}: {Promoting} {Strategies} and {Tactics} for {Engagement} of {Underrepresented} {Groups} in {Undergraduate} {Research} via {Institutional} {Diversity} {Action} {Plans}},
	volume = {19},
	issn = {1557-5284},
	url = {https://www.learntechlib.org/p/182980},
	number = {1},
	journal = {Journal of STEM Education},
	author = {Pierszalowski, Sophie and Vue, Rican and Bouwma-Gearhart, Jana},
	month = mar,
	year = {2018},
	note = {Publisher: Laboratory for Innovative Technology in Engineering Education (LITEE)},
}

@article{dolan2016,
	title = {Course-based {Undergraduate} {Research} {Experiences}: {Current} knowledge and future directions},
	volume = {1},
	url = {https://sites.nationalacademies.org/cs/groups/dbassesite/documents/webpage/dbasse_177288.pdf},
	urldate = {2022-06-21},
	journal = {Natl Res Counc Comm Pap},
	author = {Dolan, Erin L.},
	year = {2016},
}

@article{oliver2023,
  title = {Student experiences with authentic research in a remote, introductory course-based undergraduate research experience in physics},
  author = {Oliver, Kristin A. and Werth, Alexandra and Lewandowski, H. J.},
  journal = {Phys. Rev. Phys. Educ. Res.},
  volume = {19},
  issue = {1},
  pages = {010124},
  numpages = {18},
  year = {2023},
  month = {Mar},
  publisher = {American Physical Society},
  doi = {10.1103/PhysRevPhysEducRes.19.010124},
  url = {https://link.aps.org/doi/10.1103/PhysRevPhysEducRes.19.010124}
}

@article{corwin_effects_2018,
	title = {Effects of {Discovery}, {Iteration}, and {Collaboration} in {Laboratory} {Courses} on {Undergraduates}’ {Research} {Career} {Intentions} {Fully} {Mediated} by {Student} {Ownership}},
	volume = {17},
	issn = {1931-7913},
	url = {https://www.lifescied.org/doi/10.1187/cbe.17-07-0141},
	doi = {10.1187/cbe.17-07-0141},
	language = {english},
	number = {2},
	urldate = {2023-01-22},
	journal = {CBE—Life Sciences Education},
	author = {Corwin, Lisa A. and Runyon, Christopher R. and Ghanem, Eman and Sandy, Moriah and Clark, Greg and Palmer, Gregory C. and Reichler, Stuart and Rodenbusch, Stacia E. and Dolan, Erin L.},
	editor = {Hewlett, James},
	month = jun,
	year = {2018},
	pages = {ar20},
}

@misc{ep3,
  howpublished = "\url{https://ep3guide.org/}"
}

@article{russell2007,
author = {Susan H. Russell  and Mary P. Hancock  and James McCullough },
title = {Benefits of Undergraduate Research Experiences},
journal = {Science},
volume = {316},
number = {5824},
pages = {548-549},
year = {2007},
doi = {10.1126/science.1140384},
URL = {https://www.science.org/doi/abs/10.1126/science.1140384},
eprint = {https://www.science.org/doi/pdf/10.1126/science.1140384}}

@article{hanauer2017,
author = {David I. Hanauer  and Mark J. Graham  and SEA-PHAGES and Laura Betancur  and Aiyana Bobrownicki  and Steven G. Cresawn  and Rebecca A. Garlena  and Deborah Jacobs-Sera  and Nancy Kaufmann  and Welkin H. Pope  and Daniel A. Russell  and William R. Jacobs  and Viknesh Sivanathan  and David J. Asai  and Graham F. Hatfull  and Luis Actis  and Tammy Adair  and Sandra Adams  and Richard Alvey  and Kirk Anders  and Winston A. Anderson  and Lisa Antoniacci  and Mary Ayuk  and Frederick Baliraine  and Mitchell Balish  and Sarah Ball  and Brad Barbazuk  and Nazir Barekzi  and Alessandra Barrera  and Charlotte Berkes  and Aaron Best  and Suparna Bhalla  and Larry Blumer  and Dave Bollivar  and J. Alfred Bonilla  and Kim Borges  and Beckie Bortz  and Donald Breakwell  and Caroline Breitenberger  and Tim Breton  and Christopher Brey  and Jerald S. Bricker  and Laura Briggs  and Eribo Broderick  and Tessa Durham Brooks  and Victoria Brown-Kennerly  and Mike Buckholt  and Kristen Butela  and Christine Byrum  and Donna Cain  and Susan Carson  and Steve Caruso  and Laurie Caslake  and Catherine Chia  and Hui-Min Chung  and Kari Clase  and Barb Clement  and Stephanie Conant  and Bernadette Connors  and Roy Coomans  and William D'Angelo  and Tom D'Elia  and Charles J. Daniels  and Luke Daniels  and Bill Davis  and Kristi DeCourcy  and Randy DeJong  and Kristen Delaney-Nguyen  and Veronique Delesalle  and Arturo Diaz  and Leon Dickson  and Jean Doty  and Erin Doyle  and David Dunbar  and Jennifer Easterwood  and Megan Eckardt  and Nicholas Edgington  and Sarah Elgin  and Marcy Erb  and Ivan Erill  and Kayla Fast  and Christy Fillman  and Ann Findley  and Emily Fisher  and Christine Fleischacker  and Marie Fogarty  and Greg Frederick  and Victoria Frost  and Emily Furbee  and Maria Gainey  and Isaura Gallegos  and Chris Gissendanner  and Urszula Golebiewska  and Julianne Grose  and Sarah Grubb  and Nancy Guild  and Susan Gurney  and Grant Hartzog  and J. Robert Hatherill  and Charles Hauser  and Heather Hendrickson  and Christopher Herren  and John Hinz  and Eric Ho  and Sandra Hope  and Lee Hughes  and Anna Hull  and Keith Hutchison  and Sharon Isern  and Gary Janssen  and Jonathan Jarvik  and Allison Johnson  and Nancy Jones  and Jacob Kagey  and Michael Kart  and Joanna Katsanos  and Tracy Keener  and Margaret Kenna  and Rodney King  and Christina King-Smith  and Bridgette Kirkpatrick  and Karen Klyczek  and Helen Koch  and Ann Koga  and Chris Korey  and Greg Krukonis  and Bob Kurt  and Steven Leadon  and Janine LeBlanc-Straceski  and Jeremy Lee  and Julia Lee-Soety  and Lynn Lewis  and Lisa Limeri  and Joy Little  and Manuel Llano  and Javier Lopez  and Christina MacLaren  and John Makemson  and Stephanie Martin  and Dmitri Mavrodi  and Natalie McGuier  and Angela McKinney  and Jeffrey McLean  and Evan Merkhofer  and Scott Michael  and Eric Miller  and Swarna Mohan  and Sally Molloy  and Kirsten Monsen-Collar  and Denise Monti  and Alison Moyer  and Jim Neitzel  and Peter Nelson  and Rob Newman  and Byron Noordewier  and Ola Olapade  and Manuel Ospina-Giraldo  and Shallee Page  and Carleitta Paige-Anderson  and Dana Pape-Zambito  and Peter Park  and Jordan Parker  and Marisa Pedulla  and Alex Peister  and Pat Pfaffle  and Giorgia Pirino  and Marie Pizzorno  and Ruth Plymale  and Joe Pogliano  and Kit Pogliano  and Ann Powell  and Marianne Poxleitner  and Mary Preuss  and Nathan Reyna  and Jenna Rickus  and Claire Rinehart  and Courtney Robinson  and Mauricio Rodriguez-Lanetty  and Germán Rosas-Acosta  and Joe Ross  and Naomi Rowland  and David Royer  and Michael Rubin  and Rachna Sadana  and Margaret Saha  and Sanghamitra Saha  and Michael Sandel  and Tom Sasek  and Lori Saunders  and Ken Saville  and Anne Scherer  and Joel Schildbach  and Stephanie Schroeder  and J. Reid Schwebach  and Muhammad Seegulam  and Miriam Segura-Totten  and Chris Shaffer  and Ryan Shanks  and April Sipprell  and Tina Slowan-Pomeroy  and Kevin Smith  and Mary Ann Smith  and Martha Smith-Caldas  and Joyce Stamm  and Stephanie Stockwell  and Emily Stowe  and Joseph Stukey  and C. Nicole Sunnen  and Brian Tarbox  and Sarah Taylor  and Louise Temple  and Marsha Timmerman  and Deb Tobiason  and Sara Tolsma  and Melissa Torres  and Carole Twichell  and Ana Maria Valle-Rivera  and Edwin Vazquez  and Jose Villagomez  and Stephanie Voshell  and Jamie Wallen  and Rob Ward  and Vassie Ware  and Marcie Warner  and Jackie Washington  and Scott Weir  and John Wertz  and Daniel Westholm  and Kathleen Weston-Hafer  and Kristi Westover  and JoAnn Whitefleet-Smith  and Allison Wiedemeier  and Mike Wolyniak  and Wenbo Yan  and Gerard P. Zegers  and Daisy Zhang  and Ana Zimmerman },
title = {An inclusive Research Education Community (iREC): Impact of the SEA-PHAGES program on research outcomes and student learning},
journal = {Proceedings of the National Academy of Sciences},
volume = {114},
number = {51},
pages = {13531-13536},
year = {2017},
doi = {10.1073/pnas.1718188115},
URL = {https://www.pnas.org/doi/abs/10.1073/pnas.1718188115},
eprint = {https://www.pnas.org/doi/pdf/10.1073/pnas.1718188115}}

@article{mason2023,
  title={Coronal Heating as Determined by the Solar Flare Frequency Distribution Obtained by Aggregating Case Studies},
  author={Mason, James Paul and Werth, Alexandra and West, Colin G and Youngblood, Allison and Woodraska, Donald L and Peck, Courtney L and Aradhya, Arvind J and Cai, Yijian and Chaparro, David and Erikson, James W and others},
  journal={The Astrophysical Journal},
  volume={948},
  number={2},
  pages={71},
  year={2023},
  publisher={IOP Publishing}
}

@article{werth_eclass_2023,
  title = {Enhancing students' views of experimental physics through a course-based undergraduate research experience},
  author = {Werth, Alexandra and West, Colin G. and Sulaiman, Nidhal and Lewandowski, H. J.},
  journal = {Phys. Rev. Phys. Educ. Res.},
  volume = {19},
  issue = {2},
  pages = {020151},
  numpages = {13},
  year = {2023},
  month = {Oct},
  publisher = {American Physical Society},
  doi = {10.1103/PhysRevPhysEducRes.19.020151},
  url = {https://link.aps.org/doi/10.1103/PhysRevPhysEducRes.19.020151}
}

@article{cohenkappa,
author = {McHugh, Mary},
year = {2012},
month = {10},
pages = {276-82},
title = {Interrater reliability: The kappa statistic},
volume = {22},
journal = {Biochemia medica : časopis Hrvatskoga društva medicinskih biokemičara / HDMB},
doi = {10.11613/BM.2012.031}
}

@article{Walkup2020,
   author = {John R Walkup and Roger A Key and Sean Patrick Duncan and Avery E Sheldon and Michael A Walkup},
   doi = {10.1088/1361-6552/ab51fb},
   issn = {0031-9120},
   issue = {1},
   journal = {Physics Education},
   month = {1},
   pages = {015010},
   title = {Catastrophic cancellation in elastic collision lab experiments},
   volume = {55},
   url = {https://iopscience.iop.org/article/10.1088/1361-6552/ab51fb},
   year = {2020},
}

@misc{karstcurenet,
  title = {Karst Study Using Geophysics at Bracken Bat Cave Preserve},
  howpublished = {\url{https://serc.carleton.edu/curenet/institutes/misc2022/examples/277988.html}},
  note = {Accessed: 2024-04-28}
}

@misc{walkcupcurenet,
  title = {Statistics in Physics Lab: Catastrophic Cancellation},
  howpublished = {\url{https://serc.carleton.edu/curenet/collection/235539.html}},
  note = {Accessed: 2024-04-28}
}

@article{jensen2023,
    author = {Jensen, Mikkel Herholdt and Morris, Eliza J. and Ray, M. W.},
    title = "{Implementing a course-based authentic learning experience with upper- and lower-division physics classesa)}",
    journal = {American Journal of Physics},
    volume = {91},
    number = {9},
    pages = {696-700},
    year = {2023},
    month = {09},
    issn = {0002-9505},
    doi = {10.1119/5.0137141},
    url = {https://doi.org/10.1119/5.0137141}
}

@article{mrazcraig2018,
 ISSN = {0047231X, 19434898},
 URL = {https://www.jstor.org/stable/26491348},
 author = {Jennifer A. Mraz-Craig and Kristy L. Daniel and Carrie J. Bucklin and Chandrani Mishra and Laila Ali and Kari L. Clase},
 journal = {Journal of College Science Teaching},
 number = {1},
 pages = {68--75},
 publisher = {National Science Teachers Association},
 title = {Student Identities in Authentic Course-Based Undergraduate Research Experience},
 urldate = {2024-05-06},
 volume = {48},
 year = {2018}
}

@article{rodenbusch2016early,
  title={Early engagement in course-based research increases graduation rates and completion of science, engineering, and mathematics degrees},
  author={Rodenbusch, Stacia E and Hernandez, Paul R and Simmons, Sarah L and Dolan, Erin L},
  journal={CBE—Life Sciences Education},
  volume={15},
  number={2},
  pages={ar20},
  year={2016},
  publisher={Am Soc Cell Biol}
}

@article{holmes2017value,
  title={Value added or misattributed? A multi-institution study on the educational benefit of labs for reinforcing physics content},
  author={Holmes, NG and Olsen, Jack and Thomas, James L and Wieman, Carl E},
  journal={Physical Review Physics Education Research},
  volume={13},
  number={1},
  pages={010129},
  year={2017},
  publisher={APS}
}

@book{corwin2022confronting,
  title={Confronting Failure: Approaches to Building Confidence and Resilience in Undergraduate Researchers.},
  author={Corwin, Lisa A and Charkoudian, Louise K},
  year={2022},
  publisher={ERIC}
}

@incollection{qualresearch,
Author = "Valerie Otero and Danielle Harlow", Editor = "Charles Henderson and Kathleen Harper", Title = {Getting Started in Qualitative Physics Education Research}, BookTitle = {Getting Started in PER}, Publisher = {American Association of Physics Teachers}, Address = {College Park}, Volume = {2}, Edition = {1}, Month = {June}, Year = {2009} }

@inproceedings{elbenzayas2016, Author = "Melissa Eblen-Zayas", Title = {The impact of metacognitive activities on student attitudes towards experimental physics}, BookTitle = {Physics Education Research Conference 2016}, Pages = {104-107}, Address = {Sacramento, CA}, Series = {PER Conference}, Month = {July 20-21}, Year = {2016} }

@article{may_labhistory2023,
  title = {Historical analysis of innovation and research in physics instructional laboratories: Recurring themes and future directions},
  author = {May, Jason M.},
  journal = {Phys. Rev. Phys. Educ. Res.},
  volume = {19},
  issue = {2},
  pages = {020168},
  numpages = {20},
  year = {2023},
  month = {Dec},
  publisher = {American Physical Society},
  doi = {10.1103/PhysRevPhysEducRes.19.020168},
  url = {https://link.aps.org/doi/10.1103/PhysRevPhysEducRes.19.020168}
}

@book{kozminski2014aapt,
    title = {{AAPT Recommendations for the Undergraduate Physics Laboratory Curriculum}},
    year = {2014},
    author = {{AAPT Committee on Laboratories}},
    publisher = {Am Assoc Phys Teach},
    url = {https://www.aapt.org/Resources/upload/LabGuidlinesDocument_EBendorsed_nov10.pdf},
    institution = {Am Assoc Phys Teach}
}

@article{galvez2010,
    author = {Galvez, Enrique and Singh, Chandralekha},
    title = {Introduction to the Theme Issue on Experiments and Laboratories in Physics Education},
    journal = {American Journal of Physics},
    volume = {78},
    number = {5},
    pages = {453-454},
    year = {2010},
    month = {05},
    issn = {0002-9505},
    doi = {10.1119/1.3399125},
    url = {https://doi.org/10.1119/1.3399125}
}

@article{hu2017,
  title = {Qualitative investigation of students' views about experimental physics},
  author = {Hu, Dehui and Zwickl, Benjamin M. and Wilcox, Bethany R. and Lewandowski, H. J.},
  journal = {Phys. Rev. Phys. Educ. Res.},
  volume = {13},
  issue = {2},
  pages = {020134},
  numpages = {12},
  year = {2017},
  month = {Nov},
  publisher = {American Physical Society},
  doi = {10.1103/PhysRevPhysEducRes.13.020134},
  url = {https://link.aps.org/doi/10.1103/PhysRevPhysEducRes.13.020134}
}

@article{holmes2018introductory,
  title={Introductory physics labs: We can do better},
  author={Holmes, Natasha G and Wieman, Carl E},
  journal={Physics today},
  volume={71},
  number={1},
  pages={38--45},
  year={2018},
  publisher={AIP Publishing}
}

@article{wilcox2017_skilldev,
  title = {Developing skills versus reinforcing concepts in physics labs: Insight from a survey of students' beliefs about experimental physics},
  author = {Wilcox, Bethany R. and Lewandowski, H. J.},
  journal = {Phys. Rev. Phys. Educ. Res.},
  volume = {13},
  issue = {1},
  pages = {010108},
  numpages = {9},
  year = {2017},
  month = {Feb},
  publisher = {American Physical Society},
  doi = {10.1103/PhysRevPhysEducRes.13.010108},
  url = {https://link.aps.org/doi/10.1103/PhysRevPhysEducRes.13.010108}
}

@article{wilcox2016_openvguided,
  title = {Open-ended versus guided laboratory activities:Impact on students' beliefs about experimental physics},
  author = {Wilcox, Bethany R. and Lewandowski, H. J.},
  journal = {Phys. Rev. Phys. Educ. Res.},
  volume = {12},
  issue = {2},
  pages = {020132},
  numpages = {8},
  year = {2016},
  month = {Oct},
  publisher = {American Physical Society},
  doi = {10.1103/PhysRevPhysEducRes.12.020132},
  url = {https://link.aps.org/doi/10.1103/PhysRevPhysEducRes.12.020132}
}

@article{kretchmer2024,
  author = {M. Kretchmer and R. Merritt and H. Lewandowski},
  title = {Exploring student beliefs of traditional physics laboratory coursework in relation to authentic research},
  year = {2024},
  journal = {2024 PERC Proceedings},
  pages = {230},
  url = {https://www.per-central.org/items/detail.cfm?ID=16901},
  doi = {10.1119/perc.2024.pr.Kretchmer},
}

@article{merritt2024,
  author = {R. Merritt and H. Lewandowski},
  title = {Physics Instructor Views on course-based undergraduate research experiences (CUREs)},
  year = {2024},
  journal = {2024 PERC Proceedings},
  pages = {293},
  url = {https://www.per-central.org/items/detail.cfm?ID=16912},
  doi = {10.1119/perc.2024.pr.Merritt},
}

@article{geschwindtax_2024,
  title = {Development of a global landscape of undergraduate physics laboratory courses},
  author = {Geschwind, Gayle and Alemani, Micol and Fox, Michael F. J. and Logman, P. S. W. M. and Tufino, Eugenio and Lewandowski, H. J.},
  journal = {Phys. Rev. Phys. Educ. Res.},
  volume = {20},
  issue = {2},
  pages = {020117},
  numpages = {29},
  year = {2024},
  month = {Sep},
  publisher = {American Physical Society},
  doi = {10.1103/PhysRevPhysEducRes.20.020117},
  url = {https://link.aps.org/doi/10.1103/PhysRevPhysEducRes.20.020117}
}

@article{holmeslew2020,
  title = {Investigating the landscape of physics laboratory instruction across North America},
  author = {Holmes, N. G. and Lewandowski, H. J.},
  journal = {Phys. Rev. Phys. Educ. Res.},
  volume = {16},
  issue = {2},
  pages = {020162},
  numpages = {10},
  year = {2020},
  month = {Dec},
  publisher = {American Physical Society},
  doi = {10.1103/PhysRevPhysEducRes.16.020162},
  url = {https://link.aps.org/doi/10.1103/PhysRevPhysEducRes.16.020162}
}

@article{eclass,
  title = {Epistemology and expectations survey about experimental physics: Development and initial results},
  author = {Zwickl, Benjamin M. and Hirokawa, Takako and Finkelstein, Noah and Lewandowski, H. J.},
  journal = {Phys. Rev. ST Phys. Educ. Res.},
  volume = {10},
  issue = {1},
  pages = {010120},
  numpages = {14},
  year = {2014},
  month = {Jun},
  publisher = {American Physical Society},
  doi = {10.1103/PhysRevSTPER.10.010120},
  url = {https://link.aps.org/doi/10.1103/PhysRevSTPER.10.010120}
}

@article{spruce,
  title = {Survey of physics reasoning on uncertainty concepts in experiments: An assessment of measurement uncertainty for introductory physics labs},
  author = {Vignal, Michael and Geschwind, Gayle and Pollard, Benjamin and Henderson, Rachel and Caballero, Marcos D. and Lewandowski, H. J.},
  journal = {Phys. Rev. Phys. Educ. Res.},
  volume = {19},
  issue = {2},
  pages = {020139},
  numpages = {19},
  year = {2023},
  month = {Oct},
  publisher = {American Physical Society},
  doi = {10.1103/PhysRevPhysEducRes.19.020139},
  url = {https://link.aps.org/doi/10.1103/PhysRevPhysEducRes.19.020139}
}

@article{wooten2018astrocure,
  title={Investigating introductory astronomy students’ perceived impacts from participation in course-based undergraduate research experiences},
  author={Wooten, Michelle M and Coble, Kim and Puckett, Andrew W and Rector, Travis},
  journal={Physical Review Physics Education Research},
  volume={14},
  number={1},
  pages={010151},
  year={2018},
  publisher={APS}
}

@article{rabosky2025muoncure,
  title={A CURE (Course-Based Undergraduate Research) for Advanced Physics Lab},
  author={Rabosky, Kristin and Armstrong, John and Johnston, Adam},
  journal={The Physics Teacher},
  volume={63},
  number={1},
  pages={53--55},
  year={2025},
  publisher={AIP Publishing},
  doi= {10.1119/5.0186665}, 
  url = {https://pubs.aip.org/aapt/pte/article/63/1/53/3328593}
}

@article{brownell2012barriers,
  title={Barriers to faculty pedagogical change: Lack of training, time, incentives, and… tensions with professional identity?},
  author={Brownell, Sara E and Tanner, Kimberly D},
  journal={CBE—Life Sciences Education},
  volume={11},
  number={4},
  pages={339--346},
  year={2012},
  publisher={American Society for Cell Biology}
}

@article{shortlidge2016assess,
  title={How to assess your CURE: a practical guide for instructors of course-based undergraduate research experiences},
  author={Shortlidge, Erin E and Brownell, Sara E},
  journal={Journal of microbiology \& biology education},
  volume={17},
  number={3},
  pages={399--408},
  year={2016},
  publisher={American Society of Microbiology}
}

@article{orton2025challenges,
  title={Challenges for activating undergraduate research: a summary from the 2021 American Society for Microbiology Conference for Undergraduate Educators},
  author={Orton, Ginger and Barnes, Matthew A and Syed, Shifath Bin and Reid, Joshua W and Smith, Allie C},
  journal={Journal of Microbiology and Biology Education},
  pages={e00099--24},
  year={2025},
  publisher={American Society for Microbiology 1752 N St., NW, Washington, DC}
}

@article{govindan2020fear,
  title={Fear of the CURE: a beginner’s guide to overcoming barriers in creating a course-based undergraduate research experience},
  author={Govindan, Brinda and Pickett, Sarah and Riggs, Blake},
  journal={Journal of Microbiology \& Biology Education},
  volume={21},
  number={2},
  pages={50},
  year={2020},
  publisher={American Society of Microbiology}
}

@article{lopatto2014central,
  title={A central support system can facilitate implementation and sustainability of a classroom-based undergraduate research experience (CURE) in genomics},
  author={Lopatto, David and Hauser, Charles and Jones, Christopher J and Paetkau, Don and Chandrasekaran, Vidya and Dunbar, David and MacKinnon, Christy and Stamm, Joyce and Alvarez, Consuelo and Barnard, Daron and others},
  journal={CBE—Life Sciences Education},
  volume={13},
  number={4},
  pages={711--723},
  year={2014},
  publisher={American Society for Cell Biology}
}

@article{shortlidge2016faculty,
  title={Faculty perspectives on developing and teaching course-based undergraduate research experiences},
  author={Shortlidge, Erin E and Bangera, Gita and Brownell, Sara E},
  journal={BioScience},
  volume={66},
  number={1},
  pages={54--62},
  year={2016},
  publisher={Oxford University Press}
}

@article{heim2019benefits,
  title={Benefits and challenges of instructing introductory biology course-based undergraduate research experiences (CUREs) as perceived by graduate teaching assistants},
  author={Heim, Ashley B and Holt, Emily A},
  journal={CBE—Life Sciences Education},
  volume={18},
  number={3},
  pages={ar43},
  year={2019},
  publisher={American Society for Cell Biology}
}

@article{spell2014redefining,
  title={Redefining authentic research experiences in introductory biology laboratories and barriers to their implementation},
  author={Spell, Rachelle M and Guinan, Judith A and Miller, Kristen R and Beck, Christopher W},
  journal={CBE—Life Sciences Education},
  volume={13},
  number={1},
  pages={102--110},
  year={2014},
  publisher={American Society for Cell Biology}
}

@article{harrison2011classroom,
  title={Classroom-based science research at the introductory level: changes in career choices and attitude},
  author={Harrison, Melinda and Dunbar, David and Ratmansky, Lisa and Boyd, Kimberly and Lopatto, David},
  journal={CBE—Life Sciences Education},
  volume={10},
  number={3},
  pages={279--286},
  year={2011},
  publisher={American Society for Cell Biology}
}

@article{graham_persistence_2013,
author = {Mark J. Graham  and Jennifer Frederick  and Angela Byars-Winston  and Anne-Barrie Hunter  and Jo Handelsman },
title = {Increasing Persistence of College Students in STEM},
journal = {Science},
volume = {341},
number = {6153},
pages = {1455-1456},
year = {2013},
doi = {10.1126/science.1240487},
URL = {https://www.science.org/doi/abs/10.1126/science.1240487},
eprint = {https://www.science.org/doi/pdf/10.1126/science.1240487}}

@article{brownell2015high,
  title={A high-enrollment course-based undergraduate research experience improves student conceptions of scientific thinking and ability to interpret data},
  author={Brownell, Sara E and Hekmat-Scafe, Daria S and Singla, Veena and Chandler Seawell, Patricia and Conklin Imam, Jamie F and Eddy, Sarah L and Stearns, Tim and Cyert, Martha S},
  journal={CBE—Life Sciences Education},
  volume={14},
  number={2},
  pages={ar21},
  year={2015},
  publisher={American Society for Cell Biology}
}

@article{rowland2016we,
  title={Do we need to design course-based undergraduate research experiences for authenticity?},
  author={Rowland, Susan and Pedwell, Rhianna and Lawrie, Gwen and Lovie-Toon, Joseph and Hung, Yu},
  journal={CBE—Life Sciences Education},
  volume={15},
  number={4},
  pages={ar79},
  year={2016},
  publisher={American Society for Cell Biology}
}

@article{cooper2019impact,
  title={The impact of broadly relevant novel discoveries on student project ownership in a traditional lab course turned CURE},
  author={Cooper, Katelyn M and Blattman, Joseph N and Hendrix, Taija and Brownell, Sara E},
  journal={CBE—Life Sciences Education},
  volume={18},
  number={4},
  pages={ar57},
  year={2019},
  publisher={American Society for Cell Biology}
}

@article{wilczek2022catalyzing,
  title={Catalyzing the development of self-efficacy and science identity: A green organic chemistry CURE},
  author={Wilczek, Luke A and Clarke, Alannah J and Guerrero Martinez, Maria del Carmen and Morin, Jesse B},
  journal={Journal of Chemical Education},
  volume={99},
  number={12},
  pages={3878--3887},
  year={2022},
  publisher={ACS Publications}
}

@article{newell2022gains,
  title={Gains in scientific identity, scientific self-efficacy, and career intent distinguish upper-level CUREs from traditional experiences in the classroom},
  author={Newell, MiKayla J and Ulrich, Paul N},
  journal={Journal of Microbiology \& Biology Education},
  volume={23},
  number={3},
  pages={e00051--22},
  year={2022},
  publisher={American Society for Microbiology 1752 N St., NW, Washington, DC}
}

@article{corwin2022students,
  title={Students’ emotions, perceived coping, and outcomes in response to research-based challenges and failures in two sequential CUREs},
  author={Corwin, Lisa A and Ramsey, Michael E and Vance, Eric A and Woolner, Elizabeth and Maiden, Stevie and Gustafson, Nina and Harsh, Joseph A},
  journal={CBE—Life Sciences Education},
  volume={21},
  number={2},
  pages={ar23},
  year={2022},
  publisher={American Society for Cell Biology}
}

@article{gin2018students,
  title={Students who fail to achieve predefined research goals may still experience many positive outcomes as a result of CURE participation},
  author={Gin, Logan E and Rowland, Ashley A and Steinwand, Blaire and Bruno, John and Corwin, Lisa A},
  journal={CBE—Life Sciences Education},
  volume={17},
  number={4},
  pages={ar57},
  year={2018},
  publisher={American Society for Cell Biology}
}

@article{goodwin2021science,
  title={Is this science? Students’ experiences of failure make a research-based course feel authentic},
  author={Goodwin, Emma C and Anokhin, Vladimir and Gray, MacKenzie J and Zajic, Daniel E and Podrabsky, Jason E and Shortlidge, Erin E},
  journal={CBE—Life Sciences Education},
  volume={20},
  number={1},
  pages={ar10},
  year={2021},
  publisher={American Society for Cell Biology}
}

@article{schraw1995cognitive,
  title={Cognitive processes in well-defined and ill-defined problem solving},
  author={Schraw, Gregory and Dunkle, Michael E and Bendixen, Lisa D},
  journal={Applied cognitive psychology},
  volume={9},
  number={6},
  pages={523--538},
  year={1995},
  publisher={Wiley Online Library}
}

@mastersthesis{micahthesis,
  title        = {Comparing Student Experiences from a Traditional Laboratory Course to a CURE Course},
  author       = {Micah Kretchmer},
  year         = 2025,
  month        = {May},
  address      = {Boulder, CO},
  url ={https://scholar.colorado.edu/concern/undergraduate_honors_theses/d217qr38q},
  school       = {University of Colorado Boulder},
  type         = {Undergraduate Honors Thesis}
}

@article{chang2018group,
  title={When group work doesn’t work: Insights from students},
  author={Chang, Yunjeong and Brickman, Peggy},
  journal={CBE—Life Sciences Education},
  volume={17},
  number={3},
  pages={ar52},
  year={2018},
  publisher={American Society for Cell Biology}
}

@misc{cimer,
  howpublished = "\url{https://cimerproject.org/}"
}

@article{hewitt2023,
  title = {Development and assessment of a course-based undergraduate research experience for online astronomy majors},
  author = {Hewitt, Heather B. and Simon, Molly N. and Mead, Chris and Grayson, Skylar and Beall, Grace L. and Zellem, Robert T. and Tock, Kal\'ee and Pearson, Kyle A.},
  journal = {Phys. Rev. Phys. Educ. Res.},
  volume = {19},
  issue = {2},
  pages = {020156},
  numpages = {26},
  year = {2023},
  month = {Nov},
  publisher = {American Physical Society},
  doi = {10.1103/PhysRevPhysEducRes.19.020156},
  url = {https://link.aps.org/doi/10.1103/PhysRevPhysEducRes.19.020156}
}

@inproceedings{stein2018, Author = "Martin M. Stein and Emily M. Smith and Natasha G. Holmes", Title = {Confirming what we know: Understanding questionable research practices in intro physics labs}, BookTitle = {Physics Education Research Conference 2018}, Address = {Washington, DC}, Series = {PER Conference}, Month = {August 1-2}, Year = {2018} }

@inproceedings{knuth2020swx,
  title={SWx TREC's Space Weather Data Portal: a launch pad for space weather research},
  author={Knuth, Jenny and Lucas, Greg and Pankratz, Christopher K and Berger, Thomas E and Clark, Richard D and Skov, Tamitha M},
  booktitle={AGU Fall Meeting Abstracts},
  volume={2020},
  pages={SM003--0018},
  year={2020}
}

@misc{lasp_space_weather_portal,
  author       = {{Laboratory for Atmospheric and Space Physics}},
  title        = {Space Weather Data Portal},
  howpublished = {\url{https://lasp.colorado.edu/space-weather-portal/}},
  doi          = {10.25980/NMFX-XX89},

}

@article{shen2026key,
  title={Key Advancements and Emerging Trends of Perovskite Solar Cells in 2024--2025},
  author={Shen, Xiangqian and Lin, Xuesong and Su, Hongzhen and Zhang, Ziyang and Wu, Tianhao and Zhang, Jing and Peng, Yong and Zhang, Yiqiang and Zhang, Shufang and Zhou, Zhongmin and others},
  journal={Nano-Micro Letters},
  volume={18},
  number={1},
  pages={209},
  year={2026},
  publisher={Springer},
  doi = {10.1007/s40820-025-02022-6},
url = {https://link.springer.com/article/10.1007/s40820-025-02022-6},
}

@article{huperovskite2023,
title = {The Current Status and Development Trend of Perovskite Solar Cells},
journal = {Engineering},
volume = {21},
pages = {15-19},
year = {2023},
issn = {2095-8099},
doi = {10.1016/j.eng.2022.10.012},
url = {https://www.sciencedirect.com/science/article/pii/S2095809922008086},
author = {Zhelu Hu and Chenxin Ran and Hui Zhang and Lingfeng Chao and Yonghua Chen and Wei Huang}
}

@article{olasoji2025metal,
  title = {Metal halide perovskites: a platform for next-generation multifunctional devices},
 author={Olasoji, Abimbola Jacob and Park, Jin Kyoung and Lee, Hyong Joon and Song, Yunmi and Lee, David Sunghwan and Im, Sang Hyuk},
  journal = {Adv. Ind. Eng. Chem.},
  volume = {1},
  issue = {1},
  year = {2025},
  publisher = {Springer},
  doi = {10.1007/s44405-025-00011-2},
  url = {https://link.springer.com/article/10.1007/s44405-025-00011-2}
}

@article{li2018scalable,
  title={Scalable fabrication of perovskite solar cells},
  author={Li, Zhen and Klein, Talysa R and Kim, Dong Hoe and Yang, Mengjin and Berry, Joseph J and Van Hest, Maikel FAM and Zhu, Kai},
  journal={Nature Reviews Materials},
  volume={3},
  number={4},
  pages={18017},
  year={2018},
  publisher={Nature Publishing Group}
}

@article{wieman2015measuring,
  title={Measuring the impact of an instructional laboratory on the learning of introductory physics},
  author={Wieman, Carl and Holmes, Natasha G},
  journal={American Journal of Physics},
  volume={83},
  number={11},
  pages={972--978},
  year={2015},
  publisher={AIP Publishing}
}

@article{walsh2022,
  title = {Skills-focused lab instruction improves critical thinking skills and experimentation views for all students},
  author = {Walsh, Cole and Lewandowski, H. J. and Holmes, N. G.},
  journal = {Phys. Rev. Phys. Educ. Res.},
  volume = {18},
  issue = {1},
  pages = {010128},
  numpages = {18},
  year = {2022},
  month = {Apr},
  publisher = {American Physical Society},
  doi = {10.1103/PhysRevPhysEducRes.18.010128},
  url = {https://link.aps.org/doi/10.1103/PhysRevPhysEducRes.18.010128}
}

@article{etkina2010design,
  title={Design and reflection help students develop scientific abilities: Learning in introductory physics laboratories},
  author={Etkina, Eugenia and Karelina, Anna and Ruibal-Villasenor, Maria and Rosengrant, David and Jordan, Rebecca and Hmelo-Silver, Cindy E},
  journal={The Journal of the Learning Sciences},
  volume={19},
  number={1},
  pages={54--98},
  year={2010},
  publisher={Taylor \& Francis}
}

@article{deacon2011student,
  title={Student perceptions of the value of physics laboratories},
  author={Deacon, Christopher and Hajek, Allyson},
  journal={International Journal of Science Education},
  volume={33},
  number={7},
  pages={943--977},
  year={2011},
  publisher={Taylor \& Francis}
}

@article{la2021comparison,
  title={Comparison of labatorials and traditional labs: The impacts of instructional scaffolding on the student experience and conceptual understanding},
  author={La Braca, Franco and Kalman, Calvin S},
  journal={Physical Review Physics Education Research},
  volume={17},
  number={1},
  pages={010131},
  year={2021},
  publisher={APS}
}

@article{hanshawperc2015,
  author = {S. Hanshaw and Dimitri Dounas-Frazer and Heather J. Lewandowski},
  title = {Access to undergraduate research experiences at a large research university},
  year = {2015},
  journal = {2015 PERC Proceedings},
  pages = {123},
  url = {https://www.per-central.org/items/detail.cfm?ID=13852},
  doi = {10.1119/perc.2015.pr.026},
}

@article{leak2018,
  title = {Hidden factors that influence success in the optics workforce},
  author = {Leak, Anne E. and Santos, Zackary and Reiter, Erik and Zwickl, Benjamin M. and Martin, Kelly Norris},
  journal = {Phys. Rev. Phys. Educ. Res.},
  volume = {14},
  issue = {1},
  pages = {010136},
  numpages = {12},
  year = {2018},
  month = {Jun},
  publisher = {American Physical Society},
  doi = {10.1103/PhysRevPhysEducRes.14.010136},
  url = {https://link.aps.org/doi/10.1103/PhysRevPhysEducRes.14.010136}
}

@article{karimi2021strategically,
  title={Strategically addressing the soft skills gap among STEM undergraduates},
  author={Karimi, Haleh and Pina, Anthony},
  journal={Journal of Research in STEM Education},
  volume={7},
  number={1},
  pages={21--46},
  year={2021}
}

@article{alicea-munoz2021,
  title = {Transforming the preparation of physics graduate teaching assistants: Curriculum development},
  author = {Alicea-Mu\~noz, Emily and Subi\~no Sullivan, Carol and Schatz, Michael F.},
  journal = {Phys. Rev. Phys. Educ. Res.},
  volume = {17},
  issue = {2},
  pages = {020125},
  numpages = {18},
  year = {2021},
  month = {Sep},
  publisher = {American Physical Society},
  doi = {10.1103/PhysRevPhysEducRes.17.020125},
  url = {https://link.aps.org/doi/10.1103/PhysRevPhysEducRes.17.020125}
}

@article{van2011workplace,
  title={Workplace learning from a socio-cultural perspective: creating developmental space during the general practice clerkship},
  author={Van der Zwet, J and Zwietering, PJ and Teunissen, PW and Van der Vleuten, CPM and Scherpbier, AJJA},
  journal={Advances in Health Sciences Education},
  volume={16},
  number={3},
  pages={359--373},
  year={2011},
  publisher={Springer}
}

@book{lave1991situated,
  title={Situated learning: Legitimate peripheral participation},
  author={Lave, Jean and Wenger, Etienne},
  year={1991},
  publisher={Cambridge university press}
}

@article{esteban2014funds,
  title={Funds of identity: A new concept based on the funds of knowledge approach},
  author={Esteban-Guitart, Mois{\`e}s and Moll, Luis C},
  journal={Culture \& psychology},
  volume={20},
  number={1},
  pages={31--48},
  year={2014},
  publisher={Sage Publications Sage UK: London, England}
}

@article{irving_sayer2014,
  title = {Conditions for building a community of practice in an advanced physics laboratory},
  author = {Irving, Paul W. and Sayre, Eleanor C.},
  journal = {Phys. Rev. ST Phys. Educ. Res.},
  volume = {10},
  issue = {1},
  pages = {010109},
  numpages = {16},
  year = {2014}
  }

@book{vygotsky1978mind,
  title={Mind in society: The development of higher psychological processes},
  author={Vygotsky, Lev S},
  volume={86},
  year={1978},
  publisher={Harvard university press}
}

@misc{suppmat,
  note = {See Supplemental Material at [URL will be inserted by publisher] for the course overview, activity descriptions, and complete qualitative codebook, including code definitions and representative student quotations.}
}

\end{document}